\documentclass[conference, letterpaper]{IEEEtran}

\usepackage[utf8]{inputenc} 
\usepackage[T1]{fontenc}
\usepackage{url}
\usepackage{ifthen}
\usepackage[noadjust]{cite}
\usepackage[cmex10]{amsmath}
\usepackage{multirow}
\usepackage{cite}
\usepackage{amsmath,amssymb,amsfonts}
\usepackage{algorithm}
\usepackage{subcaption}
\usepackage{algpseudocode}
\usepackage{graphicx}
\usepackage{makecell}
\usepackage{textcomp}
\usepackage{xcolor}
\usepackage{soul} 
\usepackage{color,soul}
\usepackage{caption}
\usepackage{mathtools}
\usepackage{comment}
\usepackage{amsthm}
\usepackage{csquotes}
\usepackage{enumitem}
\usepackage{romannum}
\usepackage{stfloats}
\usepackage{dsfont}
\usepackage{subfiles}
\newtheorem{thm}{Theorem}
\newtheorem{lem}{Lemma}

\newtheorem{rem}{Remark}

\newtheorem{defn}{Definition}

\newtheorem{exmp}{Example}

\definecolor{mikadoyellow}{rgb}{1.0, 0.77, 0.05}   

\allowdisplaybreaks

\soulregister\cite7
\soulregister\ref7
\soulregister\pageref7

\AtBeginDocument{\pagenumbering{arabic} }
\begin{document}

\title{Placement Delivery Arrays for Coded Caching with Shared and Private Caches
}

\author{\IEEEauthorblockN{K. K. Krishnan Namboodiri, Elizabath Peter, and B. Sundar Rajan}
\IEEEauthorblockA{Department of Electrical Communication Engineering, Indian Institute of Science, Bengaluru, India \\
\{krishnank,elizabathp,bsrajan\}@iisc.ac.in
}
}

\maketitle

\begin{abstract}	
 We consider a coded caching network consisting of a server with a library of $N$ files connected to $K$ users, where each user is equipped with a dedicated cache of size $M_p$ units. In addition to that, the network consists of $\Lambda\leq K$ helper caches, each with a size $M_h$ units. Each helper cache can serve an arbitrary number of users; however, each user can access only a single helper cache. Also, we assume that the server knows the user-to-helper cache association, defined as the sets of users connected to each helper cache, during the cache placement phase. We propose a solution for the aforementioned coded caching problem by introducing a combinatorial structure called a Shared and Private Placement Delivery Array (SP-PDA). These SP-PDAs describe the helper cache placement, private cache placement, and the server transmissions in a single array. Further, we propose a novel construction of SP-PDAs using two Placement Delivery Arrays (PDAs). Interestingly, we observe that the permutations of the columns of the two chosen PDAs result in SP-PDAs with different performances. Moreover, we characterize the conditions for selecting the best column permutations of the chosen PDAs. Furthermore, the coded caching schemes resulting from SP-PDAs subsume two existing coded caching schemes as special cases. Additionally, SP-PDAs enable the construction of coded caching schemes with much smaller subpacketization numbers -subpacketization number is defined as the number of subfiles to which a file is divided- compared to the existing schemes, without paying much in terms of rate (the size of the transmission in the delivery phase).         

\end{abstract}
\section{Introduction}
With rapid advancements in technology, there is an increase in the adoption of data-hungry applications, especially entertainment services like on-demand music and video streaming and download. The high temporal variability of these on-demand services leads to internet traffic congestion during peak hours and often leaves the network resources under-utilized during off-peak hours. Coded caching, introduced in \cite{MaN}, is a technique to tackle this temporal variability by employing and exploiting the caches at the user end. Coded caching typically works in two phases, namely a placement phase and a delivery phase. The setting in \cite{MaN} (we refer to this setting as the \textit{dedicated cache setting} and the achievable scheme presented in \cite{MaN} as the \textit{MaN scheme}) consists of a central server having a library of $N$ files connected to $K$ users through an error-free broadcast link. The users are equipped with dedicated cache memories, each of size $M$ units. In the placement phase, the server fills these caches with file contents, without knowing the user demands. Once the user demands are known to the server, it makes coded broadcast transmissions in the shared link (in the delivery phase) . These coded messages are designed such that each user can decode its demanded file from the accessible cache contents and these coded messages received in the delivery phase. The objective of a coded caching scheme is to minimize \textit{rate}, which is 
defined as the size of the transmission (in the units of files) made by the server in the delivery phase. The rate achieved by the MaN scheme is $\frac{K(1-\frac{M}{N})}{1+\frac{KM}{N}}$, which is shown to be optimal  \cite{YMA,WTP} under uncoded placement, when $N\geq K$. However, to achieve this optimal rate, the MaN scheme needs to divide each file into $\binom{K}{KM/N}$ subfiles, $KM/N \in \{0,1,2,\dots,K\}$. Therefore, the subpacketization number, defined as the number of subfiles into which a file is divided, required for the MaN scheme increases exponentially with the number of users in the system. To address this issue, Yan et.al introduced a combinatorial structure called Placement Delivery Array (PDA), which resulted in coded caching schemes with small subpacketization numbers \cite{YCTC}. 

The coded caching problem in a shared cache setting was first studied in \cite{PUE}. The shared cache network model considered  in \cite{PUE} consists of $K$ cache-less users and $\Lambda\leq K$ helper caches, where multiple users share a helper cache. Each helper cache can serve an arbitrary number of users, but each user can access only a single helper cache. The scheme presented in \cite{PUE} is optimal under uncoded placement. Later, the shared cache coded caching problem was studied under secrecy constraints in \cite{MeR1,PNR2}, and showed that it is impossible to ensure secrecy in the shared cache setting, which necessitated an additional private cache with each user having a memory at least a file size. The coding schemes in both \cite{MeR1} and \cite{PNR2} used the users' private caches only to store secret keys -to ensure secrecy- by fixing the private cache sizes at unity. Recently, in \cite{PNR3}, the authors considered a setting where both helper caches and private caches are used to prefetch the file contents (there is no secrecy constraint considered in \cite{PNR3}). Another work that considered two different sets of caches is \cite{KNM}. In the hierarchical two-layer network model considered in \cite{KNM}, the server communicates to the cache-aided users via mirrors, which are equipped their own storage memories. However, the server communicates directly to the users through a broadcast channel in the setting considered in \cite{PNR3}. Moreover in \cite{KNM}, the caches at the mirrors are not accessible to the users. Also, the number of users connected to each mirror is kept the same in \cite{KNM}. Therefore, the network model considered in \cite{PNR3} is different from the hierarchical network model considered in \cite{KNM}.

In this work, we consider the network model introduced in \cite{PNR3}, where there are $K$ users and $\Lambda\leq K$ helper caches. Each user is equipped with a private cache memory of size $M_p$ (in the units of files). In addition to that a user is connected to one of the helper caches (of size $M_h$ units). We obtain coded caching schemes for this network from a special class of PDAs called Shared and Private PDAs (SP-PDAs). Different variants of PDAs were used to obtain coded caching schemes for various settings, including dedicated cache setting \cite{CJYT,CJTY,ZCJ,MiW}, shared cache setting \cite{PeR,PNR1}, multi-antenna setting \cite{NPR,YWCQC}, hierarchical coded caching setting \cite{KWC} etc.     
\subsection{Contributions}
In this work, we consider the coded caching network introduced in \cite{PNR3}, where helper caches and private caches co-exist. Throughout this paper, we assume that the helper and private cache placements are done by the server with the knowledge of user-to-cache association (the server knows which user is connected to which helper cache, a priori). The technical contributions of this work are enlisted below:
\begin{itemize}
	\item We introduce a combinatorial structure called Shared and Private Placement Delivery Array (SP-PDA), which is a special class of PDAs (Definition \ref{def:sppda}). Corresponding to any SP-PDA, we can obtain a coded caching scheme for the considered network model where each helper cache serves a set of cache-aided users (Theorem \ref{thm1}). 
	\item We propose a novel construction of SP-PDAs from two PDAs (Section \ref{construct}). In the proposed construction, we replace each entry in the first PDA by a modified version of the sub-array of the second PDA. Further, coded caching schemes from SP-PDAs subsume \textit{Scheme 2} in \cite{PNR3} as a special case, if SP-PDAs are constructed from the PDAs corresponding to the MaN scheme (Remark \ref{remQfromMaNPDAs}). Also, \textit{Scheme 1} in \cite{PNR3} can be obtained from a class of SP-PDAs (Remark \ref{scheme1}).
	\item In our proposed construction, the permutation of the columns of the chosen PDAs will affect the performance of the resulting SP-PDA\footnote{By the performance of an SP-PDA, we mean the performance of the coded caching scheme obtained from that SP-PDA.} (Section \ref{permute}). This is an interesting result, since the permutation of the columns of PDAs does not give any performance improvement in the dedicated cache setting. However, in the shared cache setting, a performance change under column permutation of PDAs was reported in \cite{PNR1}. Further, we characterize the conditions for choosing the best column permutations of the PDAs \footnote{By best permutations, we mean the permutations of the PDAs that gives a coded caching scheme with the best possible performance.} (Theorem \ref{thm3}). 
	
	\item Finally, we show that by using the PDA obtained from \textit{Construction A} in \cite{YCTC} as one of the PDAs to construct an SP-PDA (and the second PDA used is the PDA corresponding to the MaN scheme), we can reduce the subpacketization number of the coded caching scheme approximately by a factor $\Lambda t^{t-1}$ for some $t\in [\Lambda]$, compared to \textit{Scheme 2} in \cite{PNR3}, without paying much in terms of rate (Section \ref{compare}).
\end{itemize}

\subsection{Notations}
For a positive integer $m$, the set $ \left\{1,2,\hdots,m\right\}$ is denoted as $[m]$. 
 Binomial coefficients are denoted by $\binom{n}{k} $ where $\binom{n}{k} =\frac{n!}{k!(n-k)!}$ and $\binom{n}{k} $ is zero for $n<k$.
 Bold uppercase letters are used to denote arrays and matrices. Similarly, bold lower case letters are used to denote vectors. The
 columns of an $m \times n$ array (or matrix) $\mathbf{A}$ is denoted by $\mathbf{a}_1, \mathbf{a}_2,\dots, \mathbf{a}_n$. For two vectors $\mathbf{a} =(a_1,a_2,\dots,a_m)$ and $\mathbf{b} =(b_1,b_2,\dots,b_m)$, we say $\mathbf{a}\succeq \mathbf{b}$ or $\mathbf{b}\preceq \mathbf{a}$, if $a_i\geq b_i$ for every $i\in [m]$. Finally, the set of positive integers is denoted by $\mathbb{Z}_+$. 

\section{System Model}
\label{system}
 	\begin{figure}
	\captionsetup{justification = centering}
	\captionsetup{font=small,labelfont=small}
	\begin{center}
		\captionsetup{justification = centering}
		\includegraphics[width = 0.9\columnwidth]{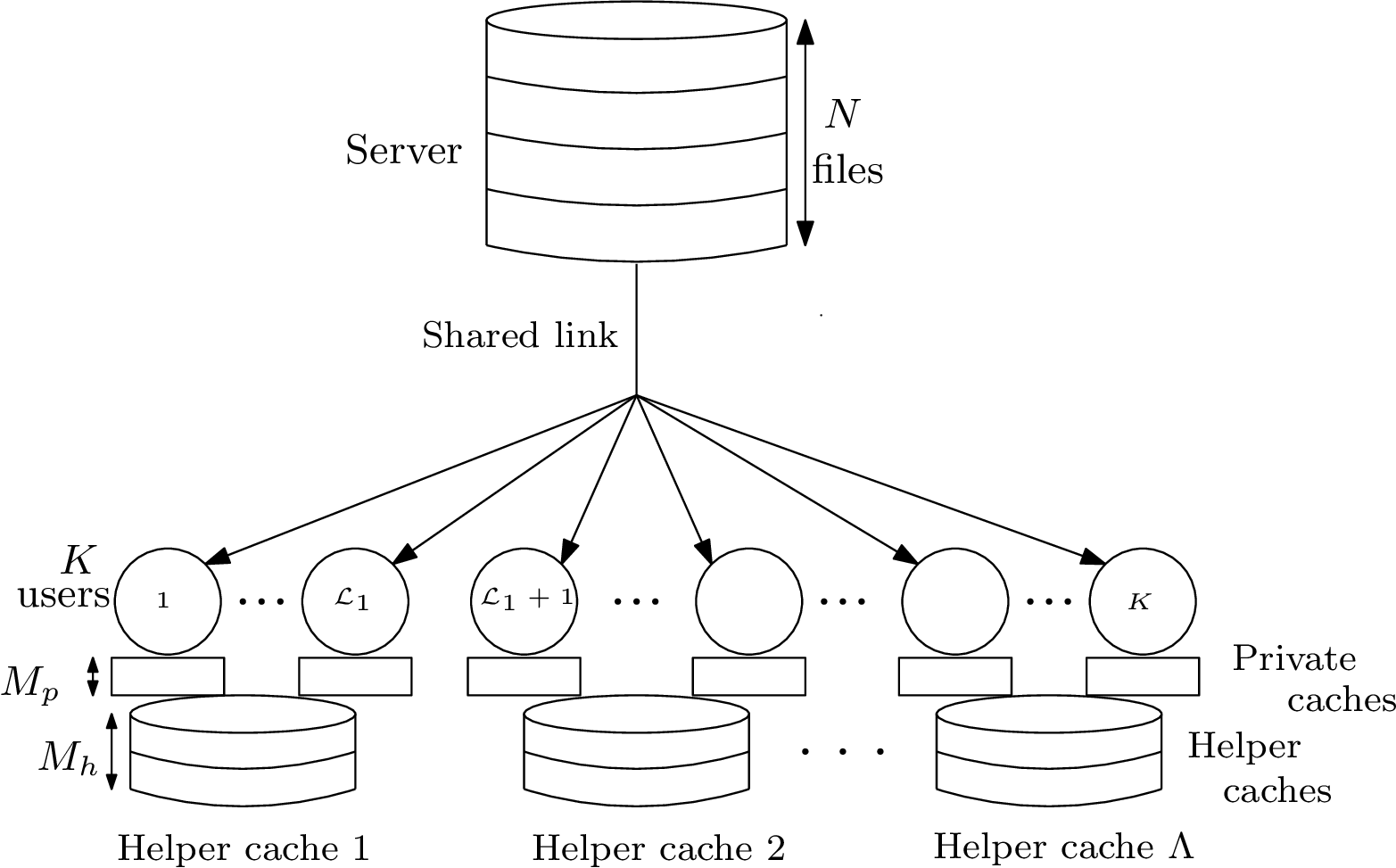}
		\caption{$(K,\Lambda,\mathcal{L},M_h.M_p,N)$ coded caching network}
		\label{network}
	\end{center}
\end{figure}
Consider a network model as shown in Fig. \ref{network}. There is a server with a library of $N$ equal-sized files $\{W_1,W_2,\ldots,W_N\}$, and is connected to $K$ users and to $\Lambda \leq K$ helper nodes. The helper nodes act as helper caches for the users, each having a size of $M_h$ files.
The users also possesses a private cache of size $M_p$ files, where $0 \leq M_h+M_p \leq N$. Each user is connected to one of the helper caches, and the association of users-to-helper caches can be arbitrary. The user-to-helper cache association  is denoted as $\mathcal{U}$, where  $\mathcal{U}=\{ \mathcal{U}_1, \ldots, \mathcal{U}_{\Lambda}\}$ and $\mathcal{U}_{\lambda}$, $\lambda \in [\Lambda]$, denotes the set of users connected to the $\lambda^{\text{th}}$ helper cache. The number of users connected to the $\lambda^{\text{th}}$ helper cache is denoted by $\mathcal{L}_{\lambda}$, and $\mathcal{L}_{\lambda}=|\mathcal{U}_{\lambda}|$. Then, $\mathcal{L}=(\mathcal{L}_1,\ldots,\mathcal{L}_{\Lambda})$ is called the association profile. Without loss of generality, we assume that the helper caches are arranged in the non-increasing order of the number of users accessing them. i.e., $\mathcal{L}_1\geq \mathcal{L}_2\geq,\dots,\geq\mathcal{L}_{\Lambda}$. 

We consider a scenario where the server knows the set of users connected to each helper cache a priori. That is, $\mathcal{U}$ is known to the server initially itself. We refer to such a network as $(K,\Lambda,\mathcal{L},M_h,M_p,N)$ coded caching network. The $(K,\Lambda,\mathcal{L},M_h,M_p,N)$ coded caching scheme operates in two phases:
\subsubsection{Placement phase}
The server places the files contents in the helper caches and the users' private caches satisfying the respective memory constraints. Let $\mathcal{C}_{\lambda}$ and $C_k$ be the contents stored in the $\lambda{\text{th}}$ helper cache and in the $k^{\text{th}}$ user's private cache, respectively. Then, $|\mathcal{C}_{\lambda}|\leq M_h$ files and $|C_k| \leq M_p$ files. The users have access to the  contents of their helper caches at zero cost. For a user $k \in [K]$, the file contents that are locally available to it is $\mathcal{C}_{\lambda_k} \cup C_k$, where $\lambda_k$ is the helper cache accessed by user $k$. To ensure full local caching gain, the placement needs to be designed such that $|\mathcal{C}_{\lambda}|= M_h$ files, $|C_k| = M_p$ files, and $\mathcal{C}_{\lambda_k} \cap C_k = \phi$. A placement policy satisfying the above conditions is plausible since $\mathcal{U}$ is known to the server at the outset. The file contents can be placed in coded or uncoded form. In this work, we consider only uncoded placement. By uncoded placement, we mean that the files are split into  subfiles and the subfiles are placed as it is in the caches without employing any coding across them. The number of subfiles constituting a file is called as subpacketization number of the scheme.

\subsubsection{Delivery Phase}
The delivery phase starts when users declare their demands to the users. Each user demands one of the $N$ files from the server. Let $d_k$ denote the file demanded by the $k^{th}$ user, and $\mathbf{d}=(d_1,\ldots,d_K)$ denotes the demands of all the users. Upon receiving the demand vector $\mathbf{d}$, the server sends transmissions over the error-free shared link to the users to satisfy their demands. Depending on the profile $\mathcal{L}$, the transmissions include both coded and uncoded messages. Using the received messages and the available cache contents, each user recovers its demanded file.

\subsubsection{Performance measure} 
The sum of sizes of all the transmissions normalized with respect to the file size is called rate, and is denoted by $R(M_s,M_p)$. Our measure of interest is the worst-case rate, hence, we consider those demand vectors $\mathbf{d} \in [N]^K$ where each $d_k, k \in [K],$ is distinct. This represents a scenario where each user demands a different file from the server. The objective 
behind any caching scheme is to appropriately design the placement and delivery policies so that the 
rate of communication over the shared link is minimized. Additionally, it is desirable to design coded caching schemes with smaller subpacketization numbers.

\section{Preliminaries}
\label{prelims}

In this section, we briefly review PDAs and see how the MaN scheme can be obtained from a class of PDAs \cite{YCTC}.

\begin{defn}[Placement Delivery Array (PDA) \cite{YCTC}]
	\label{def:pda}
	For positive integers $K, F, Z$, and $S$, an $F \times K$ array $\mathbf{P}=[p_{j,k}]$, $j \in [F]$ and $k \in [K]$, composed of a specific symbol $\star$ and $S$ positive integers $1,2,\ldots, S$, is called a $(K,F,Z,S)$ placement delivery array (PDA) if it satisfies the following three conditions: \\
	\textit{C1}. The symbol $\star$ appears $Z$ times in each column.\\
	\textit{C2}. Each integer occurs at least once in the array.\\
	\textit{C3}. For any two distinct entries $p_{j_1,k_1}$ and $p_{j_2,k_2}$, $p_{j_1,k_1}=p_{j_2,k_2}=s$ is an integer only if
	\begin{enumerate}[label=(\alph*)]
		
		\item $j_1 \neq j_2$, $k_1 \neq k_2$, i.e., they lie in distinct rows and distinct columns, and
		\item $p_{j_1,k_2}=p_{j_2,k_1}=\star$, i.e., the corresponding $2\times2$ sub-array formed by rows $j_1, j_2$ and columns $k_1,k_2$ must be of the following form:
		$\begin{bmatrix}
		s & \star\\
		\star & s
		\end{bmatrix}$
		\hspace{0.3cm}or\hspace{0.3cm}
		$\begin{bmatrix}
		\star & s \\
		s & \star
		\end{bmatrix}.$ 
		

	\end{enumerate}
\end{defn}

Every $(K,F,Z,S)$ PDA corresponds to a coded caching scheme for a dedicated cache network with parameters $K,M,$ and $N$ as in Lemma~\ref{lemm:1}.

\begin{lem}
	[\cite{YCTC}] For a given $(K, F, Z, S)$ PDA $\mathbf{P}=[p_{j,k}]_{F \times K}$, a $(K,M,N)$ coded caching scheme can be obtained with a subpacketization number $F$ and ${M}/{N}={Z}/{F}$ using \textbf{Algorithm \ref{algpda}}. For any demand vector $\mathbf{d}$, the demands of all the users are met with a rate ${S}/{F}$.
	\label{lemm:1}
\end{lem}

\begin{algorithm}
	\renewcommand{\thealgorithm}{1}
	\caption{Coded caching scheme based on PDA \cite{YCTC}}
	\label{algpda}
	\begin{algorithmic}[1]
		\Procedure{Placement}{}       
		\State Split each file $W_n$, $n \in [N]$ into $F$ subfiles: $$W_n \leftarrow\{W_{n,j}: j \in [F]\}$$
		\For{\texttt{$k \in [K]$}}
		\State  $C_k$ $\leftarrow$ $\{W_{n,j}, \forall n \in [N]$: $p_{j,k}=\star, j \in [F]\}$
		\EndFor
		\EndProcedure
		
		\Procedure{Delivery}{} 
		\For{\texttt{$s \in [S]$}}
		\State Server sends $\underset{\substack{p_{j,k}=s, j\in [F],  \textrm{\hspace{0.05cm}} k\in[K]}}{\bigoplus}W_{d_k,j}$
		\EndFor    
		\EndProcedure
	\end{algorithmic}
\end{algorithm}

In a $(K,F,Z,S)$ PDA $\mathbf{P}$, the rows represent subfiles and the columns represent users. For any $k \in [K]$, if $p_{j,k}=\star$, then it implies the $k^{\text{th}}$ user has access to the $j^{th}$ subfile of all the files. The content placed in the $k^{th}$ user's cache is denoted by $\mathcal{Z}_k$ in \textbf{Algorithm \ref{algpda}}. If $p_{j,k}=s$ is an integer, then it means that the $k^{\text{th}}$ user does not have access to the $j^{th}$ subfile of any of the files. The condition \textit{C1} guarantees that all users have access to some $Z$ subfiles of all the files. According to the delivery procedure in \textbf{Algorithm \ref{algpda}}, the server sends a linear combination of the requested subfiles indicated by the integer $s$ in the PDA. Therefore, the condition \textit{C2} implies that the number of messages transmitted by the server is $S$, and the rate is ${S}/{F}$. The condition \textit{C3} ensures decodability of the demanded files. The placement and the delivery procedures based on PDA are given in \textbf{Algorithm \ref{algpda}}.

If all the integers are appearing exactly $g$ times in a PDA, then the PDA is said to be $g$-regular.
The following lemma shows that the MaN scheme can be obtained from a regular PDA.
\begin{lem}[\cite{YCTC}]
	\label{lemmaMaNPDA}
	 For a $(K,M,N)$ caching system with ${M}/{N} \in \{0, {1}/{K}, {2}/{K}, \ldots, 1 \}$, letting $t={KM}/{N}$, there exists a $(t+1)$-regular $(K,F,Z,S)$ PDA with $F=\binom{K}{t}$, $Z=\binom{K-1}{t-1}$, and $S=\binom{K}{t+1}$.
\end{lem}
For a given $K$ and $t\in [K]$, we define the MaN PDA as the $(t+1)$-regular $(K,\binom{K}{t},\binom{K-1}{t-1},\binom{K}{t+1})$ PDA given in Lemma \ref{lemmaMaNPDA} \cite{YCTC}. The description of the MaN PDA is provided in the following paragraph.

In the MaN scheme, choose $t \in [0:K]$ such that $F = \binom{K}{t}$. Each row of the PDA is indexed by sets $\mathcal{T} \subseteq [K]$, where $|\mathcal{T}|=t$. Each user stores the subfiles $W_{n,\mathcal{T}}, \forall n \in [N],$ if $k \in \mathcal{T}$, and thus $Z=\binom{K-1}{t-1}$. In the delivery phase of the MaN scheme, there are $\binom{K}{t+1}$ transmissions. Therefore, $\binom{K}{t+1}$ distinct integers are required to represent each of those transmissions resulting in $S = \binom{K}{t+1}$. To fill the PDA with integers, a bijection $f$  is defined from the $(t+1)$-sized subsets of $\{1,2,\ldots,K\}$ to the set $\{1,2,\ldots,S\}$ such that 
\begin{equation*}
p_{\mathcal{T},k} = 
\begin{cases}
f(\mathcal{T}\cup \{k\}) & \textrm{if \hspace{0.05cm}} \mathcal{T} \not\ni k.\\
\star & \textrm{elsewhere}.
\end{cases}
\label{eq:pda_MN}
\end{equation*}

From the above expression, it can be seen that each integer appears exactly $t+1$ times. That is, the regularity of the PDA is $t+1$.

Now, we define another class of PDAs obtained from \textit{Construction A} in \cite{YCTC}.
\begin{lem}[\cite{YCTC}]
	\label{pdayctc}
	For any $q,m\in \mathbb{Z}_+$, where $q\geq 2$, there exists an $(m+1)$-regular $(q(m+1),q^m,q^{m-1},q^{m+1}-q^m)$ PDA.
\end{lem}

Next, we provide the definition of integer partition, which is helpful for our scheme description.
\begin{defn}[Integer Partition] Let $X,Y\in \mathbb{Z}_+$ such that $X\geq Y$. Then, an integer partition of $X$ of length $Y$ is a vector $(x_1,x_2,\dots,x_Y)$, where  $x_y\in\mathbb{Z}_+\cup\{0\}$ for every $y\in [Y]$, with the following properties:
	\begin{itemize}
		\item $x_1\geq x_2\geq\dots\geq x_Y$.
		\item $x_1+x_2+\dots+x_Y=X$.
	\end{itemize}
\end{defn}
\noindent For example, the vector $(4,3,2,1)$ is an integer partition of 10 of length 4, and $(5,3,2)$ is an integer partition of 10 of length 3.

 \section{Main Results}
 \label{MainResults}
 In this section, we introduce a combinatorial structure called Shared and Private Placement Delivery Array (SP-PDA). Then, we show that corresponding to any SP-PDA, we can obtain a coded caching scheme with shared and private caches. Later, we propose a construction of an SP-PDA from two PDAs. 
 \begin{defn}[Shared and Private Placement Delivery Array (SP-PDA)]
 	\label{def:sppda}
 	For positive integers $K$, $\Lambda\leq K$, $F$, $Z\leq F$, $Z^{(h)}\leq Z$, $S$, and an integer partition $\mathcal{L}=(\mathcal{L}_1,\mathcal{L}_2,\dots,\mathcal{L}_\Lambda)$ of $K$ of length $\Lambda$, an $F \times K$ array $\mathbf{Q}=[q_{j,k}]$, $j \in [F]$ and $k\in [K]$, composed of a specific symbol $\star$ and $S$ positive integers $1,2,\ldots, S$, is called a $(K,\Lambda,\mathcal{L},F,Z,Z^{(h)},S)$ shared and private placement delivery array (SP-PDA) if it satisfies the following conditions: \\
 	\textit{D1}. The array $\mathbf{Q}$ is a $(K,F,Z,S)$ PDA.\\
 	\textit{D2}. There exists a permutation $\pi$ of columns of $\mathbf{Q}$ such that $\pi((\mathbf{q}_1,\mathbf{q}_2,\dots,\mathbf{q}_K)) = (\tilde{\mathbf{q}}_1,\tilde{\mathbf{q}}_2,\dots,\tilde{\mathbf{q}}_K)$, and for a $\lambda \in [\Lambda]$, we have an array $\tilde{\mathbf{Q}}^{(\lambda)} =[\tilde{\mathbf{q}}_{1+\sum_{i=1}^{\lambda-1} \mathcal{L}_i},\tilde{\mathbf{q}}_{2+\sum_{i=1}^{\lambda-1} \mathcal{L}_i},\dots,\tilde{\mathbf{q}}_{\sum_{i=1}^{\lambda} \mathcal{L}_i}] $ of size $F\times \mathcal{L}_\lambda$. Then, for every $\lambda \in [\Lambda]$, the array $\tilde{\mathbf{Q}}^{(\lambda)}$ has at least $Z^{(h)}$ rows with only $\star$'s.
 \end{defn}
 Now, we provide an example for an SP-PDA.
 \begin{exmp}
 	The array $\mathbf{Q}$ in \eqref{Qexample} is a $(K=5, \Lambda=2, \mathcal{L}=(3,2),F=6,Z=4,Z^{(h)}=3,S=3)$ SP-PDA.
 	\begin{equation}
 		\label{Qexample}
 	\mathbf{Q}=	\begin{bmatrix}
 		\star &\star & \star &  \star  & 1\\
 		\star &\star & \star & 1& \star  \\
 		\star &\star & \star &  2  &  3  \\
 		\star &1 & 2 &  \star  &  \star  \\
 		1 &\star & 3 &  \star  &  \star \\
 		2 &3 & \star &  \star  &  \star  \\	
 	\end{bmatrix}
\end{equation}
It is easy to verify that $\mathbf{Q}$ is a $(K=5,F=6,Z=4,S=3)$ PDA. The first three columns constitute $\tilde{\mathbf{Q}}^{(1)}$ and the fourth and fifth columns constitute $\tilde{\mathbf{Q}}^{(2)}$ (under identity permutation). Note that, the first three rows of $\tilde{\mathbf{Q}}^{(1)}$ contain only $\star$'s. Similarly, rows four to six of $\tilde{\mathbf{Q}}^{(2)}$ also contain only $\star$'s.
 \end{exmp}
 \begin{rem}
 	If $\mathbf{Q}$ is a $(K,\Lambda,\mathcal{L},F,Z,Z^{(h)},S)$ SP-PDA, then $\mathbf{Q}$ is also a $(K,\Lambda,\mathcal{L},F,Z,Z^{(h)}-i,S)$ SP-PDA for every $i\in [Z^{(h)}]$. This follows from the fact that if $\mathbf{Q}$ satisfies the condition \textit{D2} for an integer $Z^{(h)}$ under certain permutation, then $\mathbf{Q}$ satisfies \textit{D2} for every $\hat{Z}^{(h)}\leq Z^{(h)}$ under the same permutation.  
 \end{rem}

 In the following theorem, we show that corresponding to any $(K,\Lambda,\mathcal{L},F,Z,Z^{(h)},S)$ SP-PDA, it is possible to obtain a $(K,\Lambda,\mathcal{L},M_h,M_p,N)$ coded caching scheme with ${M_h}/{N}={Z^{(h)}}/{F}$, and $M_p/N= (Z-Z^{(h)})/F$. 
 \begin{thm}
 	\label{thm1}
 	For a given $(K,\Lambda,\mathcal{L},F,Z,Z^{(h)},S)$ SP-PDA $\mathbf{Q}=[q_{j,k}]_{F \times K}$, a $(K,\Lambda,\mathcal{L},M_h,M_p,N)$ coded caching scheme can be obtained with subpacketization number $F$, ${M_h}/{N}={Z^{(h)}}/{F}$, and $M_p/N= (Z-Z^{(h)})/F$. For any demand vector $\mathbf{d}$, the demands of all the users can be met with rate $R={S}/{F}$.
 \end{thm}
\begin{IEEEproof}
  We show that from a $(K,\Lambda,\mathcal{L},F,Z,Z^{(h)},S)$ SP-PDA $\mathbf{Q}$, it is possible to obtain a $(K,\Lambda,\mathcal{L},M_h,M_p,N)$ coded caching scheme. Without loss of generality, we assume that $\mathbf{Q}$ satisfies \textit{D2} under identity permutation. Then, for every $\lambda\in [\Lambda]$, we define a sub-array ${\mathbf{Q}}^{(\lambda)} =[{\mathbf{q}}_{1+\sum_{i=1}^{\lambda-1} \mathcal{L}_i},{\mathbf{q}}_{2+\sum_{i=1}^{\lambda-1} \mathcal{L}_i},\dots,{\mathbf{q}}_{\sum_{i=1}^{\lambda} \mathcal{L}_i}] $ of $\mathbf{Q}$. Notice that, ${\mathbf{Q}}^{(\lambda)}$ is an array of size $F\times \mathcal{L}_\lambda$. For every $\lambda\in [\Lambda]$, we define a set
 \begin{equation*}
 	\mathcal{R}^{(\lambda)} := \{j\in [F]: \text{the $j^{\text{th}}$ row of $\mathbf{Q}^{(\lambda)}$ contains only $\star$'s}\}.
 \end{equation*}
  The condition \textit{D2} ensures that $|\mathcal{R}^{(\lambda)}|\geq Z^{(h)}$ for every $\lambda\in [\Lambda]$. Then for every $\mathcal{R}^{(\lambda)}$, we identify a subset $\mathcal{H}^{(\lambda)}$ such that $|\mathcal{H}^{(\lambda)}|=Z^{(h)}$.
  
   The coded caching scheme based on $\mathbf{Q}$ operates in two phases.
  \begin{enumerate}
  	\item \textbf{Placement Phase:} The server divides each file into $F$ subfiles. i.e., $W_n = \{W_{n,j}:j\in [F]\}$ for every $n\in [N]$.
  	
  	 First, the server fills the helper caches. The $\lambda^\text{th}$ helper cache is populated as
  	\begin{equation*}
  	\label{helpercache}\mathcal{C}_\lambda = \{W_{n,j}: j\in \mathcal{H}^{(\lambda)}, \forall n\in [N]\}.
  	\end{equation*}
  	Since $|\mathcal{H}^{(\lambda)}|=Z^{(h)}$, each helper cache contains $NZ^{(h)}$ subfiles. Each subfile has $1/F$ of a file size. Therefore, we have $M_h/N = Z^{(h)}/F$.
  	
  	Next, the server fills the private caches of the users. Consider the $k^{\text{th}}$ user, where $k\in [K]$ is such that $\sum_{i=1}^{\lambda-1} \mathcal{L}_i< k\leq  {\sum_{i=1}^{\lambda} \mathcal{L}_i} $. This means, the $k^{\text{th}}$ user is connected to the $\lambda^\text{th}$ helper cache. Then, the server populates the private cache of the $k^{\text{th}}$ user as
  	   \begin{equation*}
  	   \label{privatecache}C_k = \{W_{n,j}: q_{j,k}=\star, j\not\in \mathcal{H}^{(\lambda)}, \forall n\in [N]\}.
  	   \end{equation*}
  	   There are $Z$ $\star$'s in the $k^\text{th}$ column of $\mathbf{Q}$. However, we have $|\{j\in [F]: q_{j,k}=\star, j\not\in \mathcal{H}^{(\lambda)}\}|=Z-Z^{(h)}$. Therefore, we have $M_p/N = (Z-Z^{(h)})/F$.
  	\item \textbf{Delivery phase:} Let $\mathbf{d} =(d_1,d_2,\dots,d_K)$ be the demand vector. i.e., user $k$ demands the file $W_{d_k}$ from the server. Then for every $s\in [S]$, the server broadcasts
  	\begin{equation*}
  	\label{delivery}
  	\bigoplus_{q_{j,k}=s} W_{d_k,j}.
  	\end{equation*}
  	Note that, each coded transmission has $1/F$ of a file size. Since there are transmissions corresponding to every $s\in [S]$, the rate achieved is $S/F$.
  \end{enumerate}
The placement and the delivery phases are summarized in \textbf{Algorithm \ref{algsppda}}.

The $k^{\text{th}}$ user can access the subfiles $\{W_{n,j}: q_{j,k}=\star, \forall n\in [N]\}$ collectively from its private cache and the accessible helper cache. Since $\mathbf{Q}$ is a PDA, the decodability of the demanded files is guaranteed. That is, 
assume that the $k^{\text{th}}$ user requires $W_{d_k,j}$ from the delivery phase, then $q_{j,k}=s$ for some $s\in [S]$. From the transmission corresponding to integer $s$, the $k^{\text{th}}$ user receives,
\begin{equation*}
\label{decoding}
W_{d_k,j}\oplus\left(\bigoplus_{q_{j',k'}=s,j'\neq j,k'\neq k} W_{d_{k'},j'}\right).
\end{equation*}
However, for every $(j',k')$ such that $q_{j',k'}=s$, where $j'\neq j$ and $k'\neq k$, the condition \textit{C3} ensures that $q_{j',k}=\star$. Therefore, the $k^{\text{th}}$ user can compute $\underset{q_{j',k'}=s,j'\neq j,k'\neq k}{\bigoplus} W_{d_{k'},j'}$ using its accessible cache contents. Thus, user $k$ can decode $W_{d_k,j}$. This completes the proof of Theorem \ref{thm1}.   
\end{IEEEproof}
 
 \begin{algorithm}
 	\renewcommand{\thealgorithm}{2}
 	\caption{Coded caching scheme based on an SP-PDA}
 	\label{algsppda}
 	\begin{algorithmic}[1]
      
 		\State Split each file $W_n$, $n \in [N]$ into $F$ subfiles: $$W_n \leftarrow\{W_{n,j}: j \in [F]\}$$
 		\For{\texttt{$\lambda \in [\Lambda]$}}
 		\State Create the sub-array $${\mathbf{Q}}^{(\lambda)} \leftarrow[{\mathbf{q}}_{1+\sum_{i=1}^{\lambda-1} \mathcal{L}_i},{\mathbf{q}}_{2+\sum_{i=1}^{\lambda-1} \mathcal{L}_i},\dots,{\mathbf{q}}_{\sum_{i=1}^{\lambda} \mathcal{L}_i}] $$
 		\State Find the set $$\mathcal{R}^{(\lambda)}\leftarrow \{j\in [F]: \text{the $j^{\text{th}}$ row of $\mathbf{Q}^{(\lambda)}$ contains only $\star$'s}\}$$
 		\State Choose a subset $\mathcal{H}^{(\lambda)} \subseteq\mathcal{R}^{(\lambda)}$ such that $|\mathcal{H}^{(\lambda)}|=Z^{(h)}$
         \EndFor
         \Procedure{Helper Cache Placement}{} 
         \For{\texttt{$\lambda \in [\Lambda]$}}
 		\State  $\mathcal{C}_\lambda$ $\leftarrow$ $\{W_{n,j}: j\in \mathcal{H}^{(\lambda)}, \forall n\in [N]\}$
 		\EndFor
 		\EndProcedure
 		\Procedure{Private Cache Placement}{}       
 		\For{\texttt{$k \in [K]$}}
 		\State $\lambda\leftarrow \underset{\ell}{\arg} \sum_{i=1}^{\ell-1} \mathcal{L}_i< k\leq  {\sum_{i=1}^{\ell} \mathcal{L}_i}$
 		\State  $C_k = \{W_{n,j}: q_{j,k}\leftarrow\star, j\not\in \mathcal{H}^{(\lambda)}, \forall n\in [N]\}.$
 		\EndFor
 		\EndProcedure

 		\Procedure{Delivery}{} 
 		\For{\texttt{$s \in [S]$}}
 		\State Server sends $\underset{\substack{q_{j,k}=s j\in [F],  \textrm{\hspace{0.05cm}} k\in[K]}}{\bigoplus}W_{d_k,j}$
 		\EndFor    
 		\EndProcedure
 	\end{algorithmic}
 \end{algorithm}
 
 Next, we present an example to describe Theorem \ref{thm1}. 
 \begin{exmp}
 	Consider the $(5,2,(3,2),6,4,3,3)$ SP-PDA $\mathbf{Q}$ given in \eqref{Qexample}. From $\mathbf{Q}$, using \textbf{Algorithm \ref{algsppda}}, we obtain a $(K=5,\Lambda=2,\mathcal{L}=(3,2),M_h=\frac{N}{2},M_p=\frac{N}{6},N)$ coded caching scheme. During the placement phase, the server divides each file into 6 subfiles. For every $n\in [N]$, we have $W_n=\{W_{n,1},W_{n,1},\dots,W_{n,6}\}$. Then, the helper caches are filled as follows:
 	\begin{align*}
 		\mathcal{C}_1 &= \{W_{n,1},W_{n,2},W_{n,3}:\forall n\in [N]\},\\
 		\mathcal{C}_2 &= \{W_{n,4},W_{n,5},W_{n,6}:\forall n\in [N]\}.
 	\end{align*}
 Next, the server populates the users' private caches as
 \begin{align*}
 	C_1 &=\{W_{n,4}:\forall n\in [N]\}, C_2 =\{W_{n,5}:\forall n\in [N]\},\\C_3 &=\{W_{n,6}:\forall n\in [N]\},
 	C_4 =\{W_{n,1}:\forall n\in [N]\},\\ &\text{\hspace{1.6cm}}C_5 =\{W_{n,2}:\forall n\in [N]\}.
 \end{align*}
 Let $\mathbf{d}=(1,2,3,4,5)$ be the demand vector. Then, the server makes coded transmissions corresponding to every $s\in [3]$. Corresponding to $s=1$, the server broadcasts the coded message 
\begin{align*}
	W_{1,5}\oplus W_{2,4}\oplus W_{4,2} \oplus W_{5,1}.
\end{align*}
Similarly, corresponding to $s=2$ and $s=3$, the server broadcasts the coded messages
\begin{align*}
	&W_{1,6}\oplus W_{3,4}\oplus W_{4,3},\\
	&W_{2,6}\oplus W_{3,5}\oplus W_{5,3},
\end{align*}
respectively. Since, each of the three transmission has $1/6$ of a file size, the rate achieved is $1/2$.
 \end{exmp}

\begin{rem}
	\label{scheme1}
	The MaN PDA for $K$ and $t\in [K]$ with parameters $F=\binom{K}{t}, Z=\binom{K-1}{t-1}$, and $S=\binom{K}{t+1}$ is a $(K,\Lambda,\mathcal{L},F,Z,Z^{(h)},S)$ SP-PDA with $Z^{(h)}=\binom{K-\mathcal{L}_1}{t-\mathcal{L}_1}$, if $t\geq \mathcal{L}_1$. It holds for any column permutation of the MaN PDA. Because, if we form a sub-array of the MaN PDA by choosing any $t'\leq t$ columns, the sub-array will have exactly $\binom{K-t'}{t-t'}$ rows with only $\star$'s. The coded caching scheme presented in Theorem 1 in \cite{PNR3} (\textit{Scheme 1} in \cite{PNR3}), in fact, corresponds to this class of SP-PDAs (\textit{MaN PDAs satisfying the condition D2} for a given $K,\Lambda,$ and $\mathcal{L}$). 
\end{rem}
 
 In the following subsection, we introduce a novel construction of SP-PDAs from two PDAs. 
 \subsection{Construction of an SP-PDA from two PDAs}
 \label{construct}
 We consider a $(K,\Lambda,\mathcal{L},M_h,M_p,N)$ coded caching network, where $\mathcal{L}$ is an integer partition of $K$ of length $\Lambda$. We assume that there is a subpacketization constraint, $F\leq F_{\max}$ for some positive integer $F_{\max}$. We choose a $(\Lambda,F_1,Z_1,S_1)$ PDA $\mathbf{P}_1$ and an $(\mathcal{L}_1,F_2,Z_2,S_2)$ PDA $\mathbf{P}_2$ such that $\frac{M_h}{N} = \frac{Z_1}{F_1}$, $\frac{M_p}{N} = \frac{Z_2}{F_2}(1-\frac{Z_1}{F_1})$, and $F_1F_2\leq F_{\max}$\footnote{PDAs exist for any $K,F$ and $Z\leq F$ values, with some $S\leq K(F-Z)$.}. From $\mathbf{P}_1$ and $\mathbf{P}_2$, we construct a $(K,\Lambda,\mathcal{L},F=F_1F_2,Z=Z_1F_2+(F_1-Z_1)Z_2,Z^{(h)}=Z_1F_2,S)$ SP-PDA $\mathbf{Q}$, where $S\leq S_1S_2$ depends on both $\mathbf{P}_1$ and $\mathbf{P}_2$. The construction is as follows:
 \begin{enumerate}
 	\item Let the integer $s\in [S_1]$ appears $g_s$ times in $\mathbf{P}_1$, namely in columns $\lambda_1^{(s)}, \lambda_2^{(s)}, \dots,\lambda_{g_s}^{(s)}$, where each $\lambda_i^{(s)}\in [\Lambda]$ and $i\in [g_s]$. Without loss of generality, we assume that $\lambda_1^{(s)}<\lambda_2^{(s)}<\dots<\lambda_{g_s}^{(s)}$. For every integer $s\in [S_1]$ in $\mathbf{P}_1$, we define a function $\xi_{\mathbf{P}_1}(s)$ as the smallest index of the column in which $s$ is present, i.e., $\xi_{\mathbf{P}_1}(s) =\lambda_1^{(s)}$. Similarly, for every column index $\ell\in [\mathcal{L}_1]$ of $\mathbf{P}_2$, we define the function $\phi_{\mathbf{P}_2}(\ell)$ as the number of distinct integers present in the first $\ell$ columns (columns $1,2,\dots,\ell$) of $\mathbf{P}_2$. Then, corresponding to every integer $s\in [S_1]$ in $\mathbf{P}_1$, we construct an array $\tilde{\mathbf{P}}_2^{(s)}$ of size $F_2\times \mathcal{L}_{\xi_{\mathbf{P}_1}(s)}$, where $\tilde{\mathbf{P}}_2^{(s)}$ is obtained by deleting the columns $\mathcal{L}_{\xi_{\mathbf{P}_1}(s)}+1,\mathcal{L}_{\xi_{\mathbf{P}_1}(s)}+2,\dots,\mathcal{L}_1$ of the PDA $\mathbf{P}_2$ and replacing the $\phi_{\mathbf{P}_2}(\mathcal{L}_{\xi_{\mathbf{P}_1}(s)})$ integers in the resulting array by $1+\sum_{s'=1}^{s-1}\phi_{\mathbf{P}_2}(\mathcal{L}_{\xi_{\mathbf{P}_1}(s')}),2+\sum_{s'=1}^{s-1}\phi_{\mathbf{P}_2}(\mathcal{L}_{\xi_{\mathbf{P}_1}(s')}),\dots,\sum_{s'=1}^{s}\phi_{\mathbf{P}_2}(\mathcal{L}_{\xi_{\mathbf{P}_1}(s')})$. That is, we denote $\hat{\mathbf{P}}_2^{(s)}$ as the array obtained by deleting the columns $\mathcal{L}_{\xi_{\mathbf{P}_1}(s)}+1,\mathcal{L}_{\xi_{\mathbf{P}_1}(s)}+2,\dots,\mathcal{L}_1$ of $\mathbf{P}_2$. Then, $\hat{\mathbf{P}}_2^{(s)}$ contains $\phi_{\mathbf{P}_2}(\mathcal{L}_{\xi_{\mathbf{P}_1}(s)})$ distinct integers. Further, the smallest integer in $\hat{\mathbf{P}}_2^{(s)}$ is replaced by $1+\sum_{s'=1}^{s-1}\phi_{\mathbf{P}_2}(\mathcal{L}_{\xi_{\mathbf{P}_1}(s')})$, and the second smallest integer is replaced $2+\sum_{s'=1}^{s-1}\phi_{\mathbf{P}_2}(\mathcal{L}_{\xi_{\mathbf{P}_1}(s')})$ and so on to obtain $\tilde{\mathbf{P}}^{(s)}_2$.
 	\item To obtain the required array $\mathbf{Q}$, we replace each entry in $\mathbf{P}_1$ with an array. Specifically, each entry in the $\lambda^{\text{th}}$ column of $\mathbf{P}_1$ is replaced by an $F_2\times \mathcal{L}_\lambda$ array. Therefore, the resulting array will have a size $F_1F_2\times (\sum_{\lambda=1}^\Lambda \mathcal{L}_\lambda)$, i.e., $F\times K$. 
 	\item For a $\lambda\in [\Lambda]$, every $\star$ entry in the $\lambda^{\text{th}}$ column of $\mathbf{P}_1$ is replaced by an $F_2\times \mathcal{L}_\lambda$ array consisting of only $\star$'s. i.e., all the $F_2\mathcal{L}_{\lambda}$ entries in the array are $\star$'s. 
 	\item An integer $s$ present in $\lambda^{\text{th}}$ column of $\mathbf{P}_1$ is replaced by an array formed by deleting the $\mathcal{L}_\lambda+1,\mathcal{L}_\lambda+2,\dots, \mathcal{L}_{\xi_{\mathbf{P}_1}(s)}$ columns $\tilde{\mathbf{P}}_2^{(s)}$. Note that, $\xi_{\mathbf{P}_1}(s)$ is the smallest index of the column in which $s$ is present in $\mathbf{P}_1$, and thus $\mathcal{L}_{\xi_{\mathbf{P}_1}(s)}\geq \mathcal{L}_\lambda$. 
 	\item The total number of integers present in $\mathbf{Q}$ is $S = \left(\sum_{s=1}^{S_1}\phi_{\mathbf{P}_2}(\mathcal{L}_{\xi_{\mathbf{P}_1}(s)})\right)$.
 \end{enumerate}

\begin{rem}
	In the construction of array $\mathbf{Q}$, we have seen that $S=\left(\sum_{s=1}^{S_1}\phi_{\mathbf{P}_2}(\mathcal{L}_{\xi_{\mathbf{P}_1}(s)})\right)$. Note that, $\phi_{\mathbf{P}_2}(\mathcal{L}_{\xi_{\mathbf{P}_1}(s)})$ is the number of distinct integers present in the first $\mathcal{L}_{\xi_{\mathbf{P}_1}(s)}$ columns of $\mathbf{P}_2$. i.e., the number of distinct integers in $\tilde{\mathbf{P}}_2^{(s)}$. Therefore, we have $\phi_{\mathbf{P}_2}(\mathcal{L}_{\xi_{\mathbf{P}_1}(s)})\leq S_2$, for every $s\in [S_1]$. If, for every $s\in [S_1]$, the number of distinct integers in $\tilde{\mathbf{P}}_2^{(s)}$ is $S_2$, then $\mathbf{Q}$ will have exactly $S=S_1S_2$ integers.
\end{rem}

 From the proposed construction of SP-PDA from two PDAs, we have the following theorem.
\begin{thm}
	\label{thm2}
	For a given $K,\Lambda$, and $\mathcal{L}$, it is possible to construct a $(K,\Lambda,\mathcal{L},F,Z,Z^{(h)},S)$ SP-PDA from two PDAs, namely a $(\Lambda,F_1,Z_1,S_1)$ PDA and an $(\mathcal{L}_1,F_2,Z_2,S_2)$ PDA, where $F=F_1F_2,Z=Z_1F_2+(F_1-Z_1)Z_2,Z^{(h)}=Z_1F_2$, and $S\leq S_1S_2$. 
\end{thm}
\begin{IEEEproof}
	In Section \ref{construct}, we described the construction of an $F_1F_2\times K$ array $\mathbf{Q}$ from a $(\Lambda,F_1,Z_1,S_1)$ PDA $\mathbf{P}_1$ and an $(\mathcal{L}_1,F_2,Z_2,S_2)$ PDA $\mathbf{P}_2$. Now, we show that the array $\mathbf{Q}$ is, in fact, a $(K,\Lambda,\mathcal{L},F,Z,Z^{(h)},S)$ SP-PDA, where $F=F_1F_2,Z=Z_1F_2+(F_1-Z_1)Z_2,Z^{(h)}=Z_1F_2$, and $S\leq S_1S_2$. This means, we need to show that the array $\mathbf{Q}$ satisfies conditions \textit{D1} and \textit{D2}. The condition \textit{D1} is equivalent to conditions \textit{C1}, \textit{C2}, and \textit{C3}. 
	
	Consider the $k^{\text{th}}$ column of $\mathbf{Q}$, where $k\in [K]$. We assume that $\sum_{i=1}^{\lambda-1} \mathcal{L}_i< k\leq  {\sum_{i=1}^{\lambda} \mathcal{L}_i} $ for some $\lambda\in [\Lambda]$. Then the $k^{\text{th}}$ column of $\mathbf{Q}$ contains $F_2$ $\star$'s  corresponding to each $\star$ in the $\lambda^\text{th}$ column of $\mathbf{P}_1$. In addition to that the $k^{\text{th}}$ column of $\mathbf{Q}$ contains $Z_2$ $\star$'s corresponding to each integer in the $\lambda^\text{th}$ column of $\mathbf{P}_1$. Therefore, each column of $\mathbf{Q}$ contains $Z=Z_1F_2+(F_1-Z_1)Z_2$ $\star$'s. Thus, array $\mathbf{Q}$ satisfies \textit{C1}. Corresponding to an integer $s_1\in [S_1]$ in $\mathbf{P}_1$, array $\mathbf{Q}$ contains $\phi_{\mathbf{P}_2}(\mathcal{L}_{\xi_{\mathbf{P}_1}(s_1)})$ integers, where $\phi_{\mathbf{P}_2}(\mathcal{L}_{\xi_{\mathbf{P}_1}(s_1)})\leq S_2$. Thus, array $\mathbf{Q}$ contains $S=\sum_{s_1=1}^{S_1}\phi_{\mathbf{P}_2}(\mathcal{L}_{\xi_{\mathbf{P}_1}(s_1)})\leq S_1S_2$ distinct integers. Thus, array $\mathbf{Q}$ satisfies \textit{C2}. Next, consider an integer $s\in [S]$ in $\mathbf{Q}$. By construction, the integer $s$ in $\mathbf{Q}$ appears only in the sub-arrays corresponding to some integer $s_1$ in the PDA $\mathbf{P}_1$, where $s_1\in [S_1]$. Since, $\tilde{\mathbf{P}}_2^{(s_1)}$ satisfies the condition \textit{C3}, within a sub-array of $\mathbf{Q}$ (one sub-array corresponding to integer $s_1$ in $\mathbf{P}_1$) the integer $s$ satisfies the condition \textit{C3}. Now, assume that $q_{j_1,k_1}=q_{j_2,k_2}=s$, where $q_{j_1,k_1}$ and $q_{j_2,k_2}$ are in different sub-arrays corresponding to the integer $s_1$ in $\mathbf{P}_1=[p^{(1)}_{f,\lambda}]$, $f\in [F_1],\lambda\in [\Lambda]$. Assume that these sub-arrays are corresponding to the entries $p^{(1)}_{f_1,\lambda_1}$ $p^{(1)}_{f_2,\lambda_2}$ in $\mathbf{P}_1$, where $p^{(1)}_{f_1,\lambda_1}=p^{(1)}_{f_2,\lambda_2}=s_1$,  $f_1,f_2\in [F_1]$ and $\lambda_1,\lambda_2\in [\Lambda]$. Since $\mathbf{P}_1$ is a PDA, we have $p^{(1)}_{f_1,\lambda_2}=p^{(1)}_{f_2,\lambda_1}=\star$. Therefore, in the array $\mathbf{Q}$, we have $q_{j_1,k_2}=q_{j_2,k_1}=\star$. Thus, the array $\mathbf{Q}$ satisfies \textit{C3}, hence the array $\mathbf{Q}$ is a PDA. 
	
	An entry $\star$ in the $\lambda^\text{th}$ column of $\mathbf{P}_1$ results in $F_2$ rows with all $\star$'s in the array $\mathbf{Q}^{(\lambda)}$, where  ${\mathbf{Q}}^{(\lambda)} =[{\mathbf{q}}_{1+\sum_{i=1}^{\lambda-1} \mathcal{L}_i},{\mathbf{q}}_{2+\sum_{i=1}^{\lambda-1} \mathcal{L}_i},\dots,{\mathbf{q}}_{\sum_{i=1}^{\lambda} \mathcal{L}_i}] $. Therefore, $\mathbf{Q}^{(\lambda)}$ has at least $Z_1F_2$ rows with all $\star$'s, for every $\lambda\in [\Lambda]$. This implies, the array $\mathbf{Q}$ satisfies the condition \textit{D2} with $Z^{(h)}=Z_1F_2$ under identity permutation. Therefore, the array $\mathbf{Q}$ is a $(K,\Lambda,\mathcal{L},F,Z,Z^{(h)},S)$ SP-PDA, where $F=F_1F_2,Z=Z_1F_2+(F_1-Z_1)Z_2,Z^{(h)}=Z_1F_2$, and $S=\sum_{s_1=1}^{S_1}\phi_{\mathbf{P}_2}(\mathcal{L}_{\xi_{\mathbf{P}_1}(s_1)})\leq S_1S_2$. This completes the proof of Theorem \ref{thm2}.
\end{IEEEproof}

\begin{rem}
	If we denote our construction of SP-PDA $\mathbf{Q}$ from PDAs $\mathbf{P}_1$ and $\mathbf{P}_2$ by $\mathbf{Q}=construct(\mathbf{P}_1,\mathbf{P}_2)$, then the operation $construct(.\hspace{0.1cm},.\hspace{0.1cm})$ is not commutative. That is, in general, $construct(\mathbf{P}_1,\mathbf{P}_2)\neq construct(\mathbf{P}_2,\mathbf{P}_1)$. 
\end{rem}

In the following example, we provide the construction of a $(7,3,(4,2,1),6,4,2,5)$ SP-PDA from two PDAs.
\begin{exmp}
	From a $(3,3,1,3)$ PDA $\mathbf{P}_1$ and a $(4,2,1,2)$ PDA $\mathbf{P}_2$, we construct a $(7,3,(4,2,1),6,4,2,5)$ SP-PDA $\mathbf{Q}$ (the PDAs and the SP-PDA are given in Fig. \ref{example}). From $\mathbf{Q}$, we can obtain a $(K=7,\Lambda=3,\mathcal{L}=(4,2,1),M_h = \frac{N}{3},M_p=\frac{N}{3},N)$ coded caching scheme with a subpacketization number 6 and rate $5/6$. 
	The construction of $\mathbf{Q}$ from $\mathbf{P}_1$ and $\mathbf{P}_2$ is as follows:
	
	\begin{itemize}
		\item Since $F_2=2$ and $\mathcal{L}=(4,2,1)$, each entry in the first column of $\mathbf{P}_1$ is replaced by a $2\times 4$ array, each entry in the second column of $\mathbf{P}_1$ is replaced by a $2\times 2$ array, and each entry in the third column of $\mathbf{P}_1$ is replaced by a $2\times 1$ array.  
		\item Corresponding to an integer $s$ in $\mathbf{P}_1$, we construct an $F_2\times \mathcal{L}_{\xi_{\mathbf{P}_1}(s)}$, where $\xi_{\mathbf{P}_1}(s)\in [3]$, is the smallest index of the column (of $\mathbf{P}_1$) in which integer $s$ appears. In this example, we have $\xi_{\mathbf{P}_1}(1)=1,\xi_{\mathbf{P}_1}(2)=1$, and $\xi_{\mathbf{P}_1}(3)=2$. 
		\item Corresponding to integer $1$, we have the array
		\begin{equation*}
		\tilde{\mathbf{P}}_2^{(1)}=	\begin{bmatrix}
		\star &1 & \star & 2 \\
		1 &\star & 2 & \star  
		\end{bmatrix}.
		\end{equation*}
		Similarly, corresponding to integers $2$ and $3$, we have the arrays
		\begin{equation*}
		\tilde{\mathbf{P}}_2^{(2)}=	\begin{bmatrix}
		\star &3 & \star & 4 \\
		3 &\star & 4 & \star  
		\end{bmatrix}\text{ and }
		\tilde{\mathbf{P}}_2^{(3)}=	\begin{bmatrix}
		\star &5 \\
		5 &\star  
		\end{bmatrix}.
		\end{equation*}
		respectively.
		\item Finally, integer 1 in the first column of $\mathbf{P}_1$ is replaced by $\tilde{\mathbf{P}}_2^{(1)}$. Since $\mathcal{L}_2=2$, integer 1 in the second column of $\mathbf{P}_1$ is replaced by the sub-array formed by the first two columns of $\tilde{\mathbf{P}}_2^{(1)}$. Similarly, integer 2 in the first column of $\mathbf{P}_1$ is replaced by $\tilde{\mathbf{P}}_2^{(2)}$ and integer 2 in the third column of $\mathbf{P}_1$ is replaced by the first column of $\tilde{\mathbf{P}}_2^{(2)}$ (note that, $\mathcal{L}_3=1$). Similarly, integer 3 in the second and third columns of $\mathbf{P}_1$ is replaced by $\tilde{\mathbf{P}}_2^{(3)}$ and the first column of $\tilde{\mathbf{P}}_2^{(3)}$, respectively. 
				\begin{figure}[h]
			\captionsetup{justification = centering}
			\captionsetup{font=small,labelfont=small}
			\begin{center}
				\captionsetup{justification = centering}
				\includegraphics[width = \columnwidth]{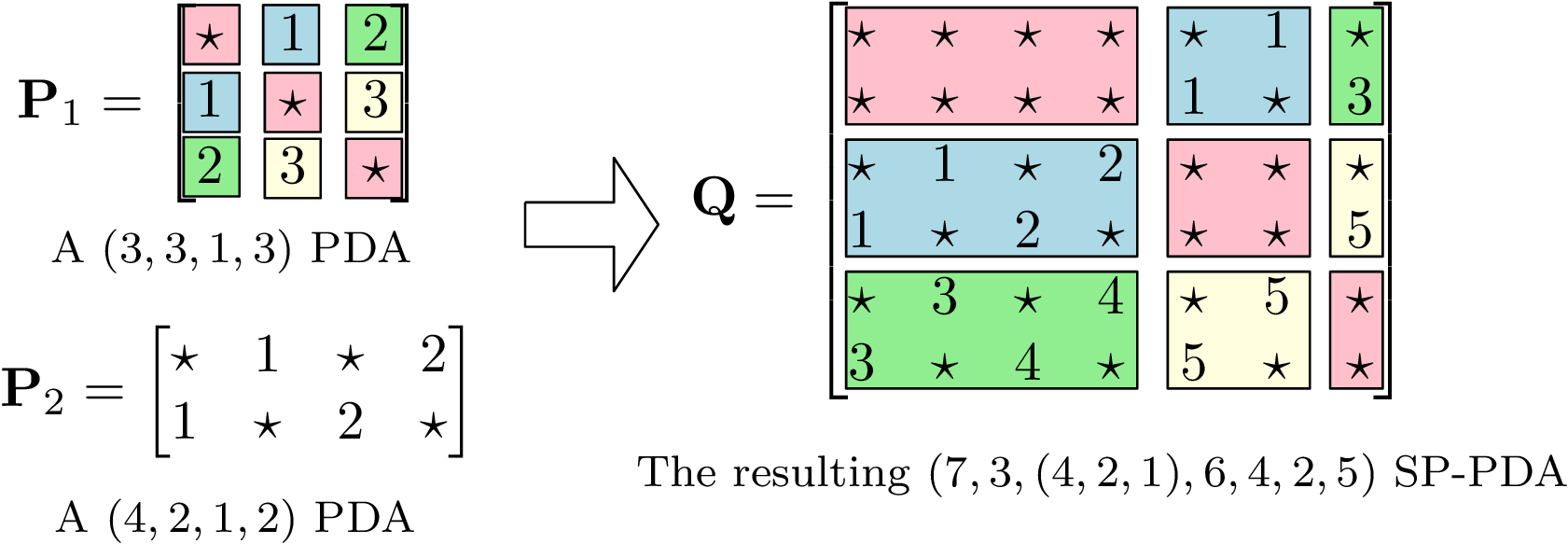}
				\caption{An illustration of constructing $\mathbf{Q}$ from $\mathbf{P}_1$ and $\mathbf{P}_2$}
				\label{example}
			\end{center}
		\end{figure}
	\end{itemize}
\end{exmp}

\begin{rem}
	\label{remQfromMaNPDAs}
	Consider the $(K,\Lambda,\mathcal{L},M_h,M_p,N)$ coded caching scheme with $M_h+M_p<N$. Let $t_1=\frac{\Lambda M_h}{N}\in [\Lambda]$ and $t_2=\frac{\mathcal{L}_1 M_p}{N(1-\frac{t_1}{\Lambda})}\in[\mathcal{L}_1]$ are non-negative integers. Then, if we choose $\mathbf{P}_1$ as the $(\Lambda,\binom{\Lambda}{t_1},\binom{\Lambda-1}{t_1-1},\binom{\Lambda}{t_1+1})$ MaN PDA and $\mathbf{P}_2$ as the $(\mathcal{L}_1,\binom{\mathcal{L}_1}{t_2},\binom{\mathcal{L}_1-1}{t_2-1},\binom{\mathcal{L}_1}{t_2+1})$ MaN PDA, the SP-PDA $\mathbf{Q}$ resulting from our proposed construction has $F=\binom{\Lambda}{t_1}\binom{\mathcal{L}_1}{t_2}, Z=\binom{\Lambda-1}{t_1-1}\binom{\mathcal{L}_1}{t_2}+\binom{\Lambda-1}{t_1}\binom{\mathcal{L}_1-1}{t_2-1},Z^{(h)}=\binom{\Lambda-1}{t_1-1}\binom{\mathcal{L}_1}{t_2}$, and $S=\sum_{n=1}^{\Lambda-t_1}\binom{\Lambda-n}{t_1}\left[\binom{\mathcal{L}_1}{t_2+1}-\binom{\mathcal{L}_1-\mathcal{L}_n}{t_2+1}\right]$ (calculation of $S$ is shown in Appendix \ref{app1}). This SP-PDA $\mathbf{Q}$ gives the required $(K,\Lambda,\mathcal{L},M_h,M_p,N)$ coded caching scheme with rate $R = \frac{\sum_{n=1}^{\Lambda-t_1}\binom{\Lambda-n}{t_1}\left[\binom{\mathcal{L}_1}{t_2+1}-\binom{\mathcal{L}_1-\mathcal{L}_n}{t_2+1}\right]}{\binom{\Lambda}{t_1}\binom{\mathcal{L}_1}{t_2}}$. This coded caching scheme is, in fact, \textit{Scheme 2} in \cite{PNR3} (Theorem 2 in \cite{PNR3}). Therefore, \textit{Scheme 2} in \cite{PNR3} is a special case of the coded caching schemes obtained from a certain class SP-PDAs, where these class of SP-PDAs is constructed with $\mathbf{P}_1$ and $\mathbf{P}_2$ being MaN PDAs.
\end{rem}
\subsection{Permuting columns to improve the performance}
\label{permute}
In this section, with the help of an example, we show that column permutations of the PDAs we begin with result in SP-PDAs with different $S$ values, which, in turn, lead to coded caching schemes with different performances (all other parameters, except $S$, will remain unchanged). We then establish a criterion for choosing the permutations of columns of the PDAs that result in an SP-PDA with the least possible $S$.

\begin{lem}
	If we apply any permutation to the columns of a $(K,F,Z,S)$ PDA, the resulting array will also be a $(K,F,Z,S)$ PDA.
\end{lem}
\begin{IEEEproof}
	The conditions \textit{C1}, \textit{C2}, and \textit{C3} are independent of column permutations. Therefore, the array resulting by applying any column permutation to a PDA is also a PDA with the same parameters.
\end{IEEEproof}

The PDAs that differ just by a column permutation are called equivalent PDAs \cite{PNR1}.
\begin{defn}[Equivalent PDAs \cite{PNR1}]
	Two PDAs $\mathbf{P}$ and $\mathbf{P}'$ are said to be equivalent, if $\mathbf{P}'$ can be obtained by permuting the columns of $\mathbf{P}$.
\end{defn}

In the following example, we show that equivalent PDAs will lead to SP-PDAs with different $S$ values.
\begin{exmp}
	Consider a coded caching network with $K=14,\Lambda=6,\mathcal{L} = (6,3,2,1,1,1), M_h =N/2$, and $M_p=N/6$. We obtain an achievable scheme for these parameters from an SP-PDA $\mathbf{Q}$ constructed from PDAs $\mathbf{P}_1$ and $\mathbf{P}_2$, where
	\begin{equation*}
\mathbf{P}_1=	\begin{bmatrix}
	\star& 3 & \star & 2 & \star& 1\\
	1 & \star & \star  & 4 & 3 & \star\\
	\star &4 & 1& \star & 2 & \star \\
	2 & \star &3 & \star & \star &4\\
	\end{bmatrix},
	\mathbf{P}_2=\begin{bmatrix}
	\star& 3 & 5 & \star& 1 & 2\\
	1 & \star & 6  & 3 & \star & 4 \\
	2 &4 & \star & 5 &6 & \star \\
	\end{bmatrix}.
	\end{equation*} 
	The SP-PDA $\mathbf{Q}$ resulting from $\mathbf{P}_1$ and $\mathbf{P}_2$ is given in \eqref{qnoperm}.
\begin{figure*}
	\begin{equation}
	\label{qnoperm}
	\mathbf{Q}=\left[\begin{array}{cccccc|ccc|cc|c|c|c}
	\star &\star &\star &\star&\star&\star &\star& 15 & 17&\star &\star  &\star&\star &\star   \\
	\star &\star &\star &\star&\star&\star & 13  & \star & 18 &\star &\star & 7&\star &1 \\
	\star &\star &\star &\star&\star&\star & 14 &16 & \star &\star &\star & 8&\star &2\\
	\hline
	\star& 3 & 5 & \star& 1 & 2  &\star &\star &\star &\star &\star &\star &\star  &\star \\
	1 & \star & 6  & 3 & \star & 4  &\star &\star &\star &\star &\star &19 &13&\star \\
	2 &4 & \star & 5 &6 & \star  &\star &\star &\star &\star&\star  &20&14&\star \\\hline
	\star &\star &\star &\star&\star&\star &\star& 21 & 23 &\star  &3 &\star &\star&\star  \\
	\star &\star &\star &\star&\star&\star & 19 & \star & 24  & 1 &\star &\star&7 &\star \\
	\star &\star &\star &\star&\star&\star & 20 &22 & \star &2 &4&\star &8&\star \\\hline
	\star& 9 & 11 & \star& 7 & 8  &\star&\star&\star &\star   &15&\star &\star&\star  \\
	7 & \star & 12  & 9 & \star & 10 &\star&\star&\star &  13&\star &\star&\star &19 \\
	8 &10 & \star & 11 &12 & \star  &\star&\star&\star & 14 &16&\star&\star &20 \\
	\end{array}\right].
	\end{equation} 
\end{figure*}
Notice that, there are 24 distinct integers in $\mathbf{Q}$. Therefore, the coded caching scheme obtained from $\mathbf{Q}$ has a rate $R=24/12=2$.
We now apply certain permutations on the columns of $\mathbf{P}_1$ and $\mathbf{P}_2$ to obtain $\mathbf{P}_1'$ and $\mathbf{P}_2'$, respectively. We have,
\begin{equation*}
\mathbf{P}_1'=	\begin{bmatrix}
\star& \star & \star & 1 & 2& 3\\
1 & \star  & 3 & \star  & 4 & \star\\
\star & 1  & 2 &\star &\star & 4 \\
2 & 3 & \star  &4 & \star & \star \\
\end{bmatrix},
\mathbf{P}_2'=\begin{bmatrix}
\star& 1 & 2 & 5& 3 & \star\\
1 & \star & 4  & 6 & \star & 3 \\
2 &6 & \star & \star & 4 &5 \\
\end{bmatrix}.
\end{equation*} 
The SP-PDA $\mathbf{Q}'$ resulting from $\mathbf{P}_1'$ and $\mathbf{P}_2'$ is given in \eqref{qwithperm}.
	\begin{figure*}
	\begin{equation}
	\label{qwithperm}
	\mathbf{Q}'=\left[\begin{array}{cccccc|ccc|cc|c|c|c}
	\star &\star &\star &\star&\star&\star &\star& \star & \star&\star &\star  &\star&\star &\star   \\
	\star &\star &\star &\star&\star&\star & \star  & \star &\star &\star &\star & 1&7&13 \\
	\star &\star &\star &\star&\star&\star & \star & \star & \star &\star &\star & 2&8 &14\\\hline
	\star& 1 & 2 & 5 & 3 & \star  &\star &\star &\star &\star &13 &\star &\star  &\star \\
	1 & \star &4  & 6 & \star & 3  &\star &\star &\star &13 &\star &\star &17&\star \\
	2 &6 & \star & \star &4 & 5  &\star &\star &\star &14&16 &\star&18&\star \\\hline
	\star &\star &\star &\star&\star&\star &\star& 1 & 2 &\star  &7 &\star &\star&\star  \\
	\star &\star &\star &\star&\star&\star & 1 & \star & 4  & 7 &\star &\star&\star &17 \\
	\star &\star &\star &\star&\star&\star & 2 & 6 & \star &8 &12&\star &\star&18 \\\hline
	\star& 7 & 8 & 11& 9 & \star &\star&13&14 &\star   &\star&\star &\star&\star  \\
	7 & \star & 10  & 12 & \star & 9 &13&\star&15&  \star&\star &17&\star &\star \\
	8 &12 & \star & \star &10 & 11 &14&16&\star & \star &\star&18&\star &\star \\
	\end{array}\right].
	\end{equation} 
\end{figure*}
The SP-PDAs $\mathbf{Q}$ and $\mathbf{Q}'$ have the same parameters, except the number of integers, $S$ in them. The number of integers in $\mathbf{Q}'$ is 18, whereas 24 integers are present in $\mathbf{Q}$. Therefore, the resulting coded caching scheme from $\mathbf{Q}'$ has a rate $R=18/12=1.5$. This reduction in rate is achieved solely by permuting the columns of the initial PDAs. Further, assume that some permutation $\pi_1$ is applied on the columns of $\mathbf{P}_1$ and some permutation $\pi_2$ is applied on the columns of $\mathbf{P}_2$ to obtain $\mathbf{P}_1^{(\pi_1)}$ and $\mathbf{P}_2^{(\pi_2)}$, respectively. Let $\mathbf{Q}^{(\pi)}$ denote the SP-PDA resulting from $\mathbf{P}_1^{(\pi_1)}$ and $\mathbf{P}_2^{(\pi_2)}$ using our construction. It is easy to verify that the SP-PDA $\mathbf{Q}^{(\pi)}$ has parameters $K=14,\Lambda=6,\mathcal{L}=(6,3,2,1,1,1),Z=8,Z^{(h)}=2$ and $S=S^{(\pi)}$, where $18\leq S^{(\pi)}\leq 24$, for any two permutations $\pi_1$ and $\pi_2$.

\end{exmp}

The following theorem gives the conditions to choose the PDAs,  from the sets of equivalent PDAs, that minimize the parameter $S$ in the resulting SP-PDA. 
\begin{thm}
	\label{thm3}
	For a $(K,\Lambda,\mathcal{L},F,Z,Z^{(h)},S)$ SP-PDA $\mathbf{Q}$ obtained from a $(\Lambda,F_1,Z_1,S_1)$ PDA $\mathbf{P}_1$ and an $(\mathcal{L}_1,F_2,Z_2,S_2)$ PDA $\mathbf{P}_2$ using our proposed construction will have the least possible $S$ if the following conditions are met:\\
		 \textit{E1}. For the PDA $\mathbf{P}_1$ and for any equivalent PDA $\mathbf{P}_1'$ of $\mathbf{P}_1$, we have $\left(\phi_{\mathbf{P}_1}(1),\phi_{\mathbf{P}_1}(2),\dots,\phi_{\mathbf{P}_1}(\Lambda)\right)\preccurlyeq (\phi_{\mathbf{P}_1'}(1),\phi_{\mathbf{P}_1'}(2),\dots,\phi_{\mathbf{P}_1'}(\Lambda))$.\\
		\textit{E2}. For the PDA $\mathbf{P}_2$ and for any equivalent PDA $\mathbf{P}_2'$ of $\mathbf{P}_2$, we have $\left(\phi_{\mathbf{P}_2}(\mathcal{L}_\Lambda),\phi_{\mathbf{P}_2}(\mathcal{L}_{\Lambda-1}),\dots,\phi_{\mathbf{P}_2}(\mathcal{L}_1)\right)\preccurlyeq \left(\phi_{\mathbf{P}_2'}(\mathcal{L}_\Lambda),\phi_{\mathbf{P}_2'}(\mathcal{L}_{\Lambda-1}),\dots,\phi_{\mathbf{P}_2'}(\mathcal{L}_1)\right)$.
\end{thm}
\begin{IEEEproof}
	First, we show that if PDA $\mathbf{P}_1$ satisfies \textit{E1}, then for any equivalent PDA $\mathbf{P}_1'$ and for a fixed PDA $\mathbf{P}_2$, the resulting SP-PDA from our construction will have at least as many integers as in an SP-PDA resulting from $\mathbf{P}_1$ and $\mathbf{P}_2$.		
		
	Assume that our construction of SP-PDA from PDAs $\mathbf{P}_1$ and $\mathbf{P}_2$ results in a $(K,\Lambda,\mathcal{L},F,Z,Z^{(h)},S)$ SP-PDA $\mathbf{Q}$. 
	Also, assume that $\mathbf{P}_1$ satisfies \textit{E1}. This means, for any equivalent PDA $\mathbf{P}_1'$ of $\mathbf{P}_1$, we have $\left(\phi_{\mathbf{P}_1'}(1),\phi_{\mathbf{P}_1'}(2),\dots,\phi_{\mathbf{P}_1'}(\Lambda)\right)\succeq \left(\phi_{\mathbf{P}_1}(1),\phi_{\mathbf{P}_1}(2),\dots,\phi_{\mathbf{P}_1}(\Lambda)\right) $. Consider an equivalent PDA $\mathbf{P}_1'$ of $\mathbf{P}_1$. Our construction of SP-PDA from PDAs $\mathbf{P}_1'$ and $\mathbf{P}_2$ results in a $(K,\Lambda,\mathcal{L},F,Z,Z^{(h)},S')$ SP-PDA $\mathbf{Q}'$. Then, we have $S = \left(\sum_{s=1}^{S_1}\phi_{\mathbf{P}_2}(\mathcal{L}_{\xi_{\mathbf{P}_1}(s)})\right)$ and $S' = \left(\sum_{s=1}^{S_1}\phi_{\mathbf{P}_2}(\mathcal{L}_{\xi_{\mathbf{P}_1'}(s)})\right)$. Now, we sort the entries in the vectors $\left(\xi_{\mathbf{P}_1}(1),\xi_{\mathbf{P}_1}(2),\dots,\xi_{\mathbf{P}_1}(S_1)\right)$ and $\left(\xi_{\mathbf{P}_1'}(1),\xi_{\mathbf{P}_1'}(2),\dots,\xi_{\mathbf{P}_1'}(S_1)\right)$ in the non-decreasing orders to obtain $\mathcal{O}$ and $\mathcal{O}'$, respectively. These sorted vectors $\mathcal{O}$ and $\mathcal{O}'$ represent the smallest indices of the columns in which the integers in $[S_1]$ appear in the PDAs $\mathbf{P}_1$ and $\mathbf{P}_1'$, respectively. Then, we have $\mathcal{O}\succeq\mathcal{O}'$, since $\phi_{\mathbf{P}_1}(\lambda)\leq\phi_{\mathbf{P}_1'}(\lambda)$ for every $\lambda\in [\Lambda]$. This means, $\mathcal{L}_{\mathcal{O}}\preceq\mathcal{L}_{\mathcal{O}'}$, where $\mathcal{L}_{\mathcal{O}}$ represents the vector obtained after sorting (in the non-decreasing order) $\left(\mathcal{L}_{\xi_{\mathbf{P}_1}(1)},\mathcal{L}_{\xi_{\mathbf{P}_1}(2)},\dots,\mathcal{L}_{\xi_{\mathbf{P}_1}(S_1)}\right)$ and $\mathcal{L}_{\mathcal{O}'}$ represents the vector obtained after sorting (in the non-decreasing order) $\left(\mathcal{L}_{\xi_{\mathbf{P}_1'}(1)},\mathcal{L}_{\xi_{\mathbf{P}_1'}(2)},\dots,\mathcal{L}_{\xi_{\mathbf{P}_1'}(S_1)}\right)$. Then, we get
	\begin{align*}
	\sum_{s=1}^{S_1}\phi_{\mathbf{P}_2}(\mathcal{L}_{\xi_{\mathbf{P}_1}(s)})&\leq\sum_{s=1}^{S_1}\phi_{\mathbf{P}_2}(\mathcal{L}_{\xi_{\mathbf{P}_1'}(s)})\\\implies S&\leq S'.
	\end{align*}  
	Next, we validate condition \textit{E2}. In order to do that, we assume that the PDA $\mathbf{P}_2$ satisfies \textit{E2}. We also assume that our construction of SP-PDA from PDAs $\mathbf{P}_1$ and $\mathbf{P}_2$ results in a $(K,\Lambda,\mathcal{L},F,Z,Z^{(h)},S)$ SP-PDA $\mathbf{Q}$. Now, consider an equivalent PDA $\mathbf{P}_2'$ of $\mathbf{P}_2$. Our construction of SP-PDA from PDAs $\mathbf{P}_1$ and $\mathbf{P}_2'$ results in a $(K,\Lambda,\mathcal{L},F,Z,Z^{(h)},S'')$ SP-PDA $\mathbf{Q}''$. Then, we have $S = \left(\sum_{s=1}^{S_1}\phi_{\mathbf{P}_2}(\mathcal{L}_{\xi_{\mathbf{P}_1}(s)})\right)$ and $S'' = \left(\sum_{s=1}^{S_1}\phi_{\mathbf{P}_2'}(\mathcal{L}_{\xi_{\mathbf{P}_1}(s)})\right)$. Since $\mathbf{P}_2$ satisfies \textit{E2}, we have $\left(\phi_{\mathbf{P}_2}(\mathcal{L}_\Lambda),\phi_{\mathbf{P}_2}(\mathcal{L}_{\Lambda-1}),\dots,\phi_{\mathbf{P}_2}(\mathcal{L}_1)\right)\preccurlyeq \left(\phi_{\mathbf{P}_2'}(\mathcal{L}_\Lambda),\phi_{\mathbf{P}_2'}(\mathcal{L}_{\Lambda-1}),\dots,\phi_{\mathbf{P}_2'}(\mathcal{L}_1)\right)$. That is, for any $\lambda\in [\Lambda]$, we have $\phi_{\mathbf{P}_2}(\mathcal{L}_\lambda)\leq \phi_{\mathbf{P}_2'}(\mathcal{L}_\lambda)$. Therefore, we get
	\begin{align*}
	\sum_{s=1}^{S_1}\phi_{\mathbf{P}_2}(\mathcal{L}_{\xi_{\mathbf{P}_1}(s)})&\leq\sum_{s=1}^{S_1}\phi_{\mathbf{P}_2'}(\mathcal{L}_{\xi_{\mathbf{P}_1}(s)})\\\implies S&\leq S''.
	\end{align*} 
	This completes the proof of Theorem \ref{thm3}.
\end{IEEEproof}

\section{Performance Comparison}
\label{compare}
In this section, we compare the performance of the coded caching schemes resulting from SP-PDAs and the performance of \textit{Scheme 2} in \cite{PNR3} (to the best of our knowledge, \cite{PNR3} is the only other work that considers the coded caching network with shared caches and private caches). We have already seen that (in Remark \ref{remQfromMaNPDAs}) \textit{Scheme 2} in \cite{PNR3} can be obtained from SP-PDAs constructed using two MaN PDAs. In order to understand the benefit that we achieve in terms of subpacketization number, we construct two SP-PDAs for a given $K,\Lambda$, and $\mathcal{L}$, namely $\mathbf{Q}_A$ and $\mathbf{Q}_{\text{MaN}}$. The SP-PDA $\mathbf{Q}_A$ is constructed by choosing $\mathbf{P}_1$ as a PDA obtained by \textit{Construction A} in \cite{YCTC}, and $\mathbf{P}_2$ as a MaN PDA, and the SP-PDA $\mathbf{Q}_{\text{MaN}}$ is constructed from two MaN PDAs. Note that, in both the cases, the second PDA $\mathbf{P}_2$ is a MaN PDA. 

Let $\mathbf{P}_1$ be the $(q(m+1),q^m,q^{m-1},q^m(q-1))$ PDA obtained from \textit{Construction A} in \cite{YCTC}. Also, let $\mathbf{P}_2$ be the $(\mathcal{L}_1,\binom{\mathcal{L}_1}{t_2},\binom{\mathcal{L}_1-1}{t_2-1},\binom{\mathcal{L}_1}{t_2+1})$ MaN PDA for some $t_2\in [\mathcal{L}_1]$. The $(K,\Lambda,\mathcal{L},F_A,Z_A,Z_A^{(h)},S_A)$ SP-PDA $\mathbf{Q}_A$ obtained from $\mathbf{P}_1$ and $\mathbf{P}_2$ using our construction has the parameters $F_A=q^m\binom{\mathcal{L}_1}{t_2},Z_A=q^{m-1}[\binom{\mathcal{L}_1}{t_2}+(q-1)\binom{\mathcal{L}_1-1}{t_2-1}], Z_A^{(h)}=q^{m-1}\binom{\mathcal{L}_1}{t_2} ,$ and $S_A\leq q^{m-1}(q-1)\left(\sum_{n=1}^{q} \binom{\mathcal{L}_1}{t_2+1}-\binom{\mathcal{L}_1-\mathcal{L}_n}{t_2+1}\right)$ (the calculation of $S_A$ is given in Appendix \ref{app2}). Since it is difficult to find the best column ordering of $\mathbf{P}_1$, we consider $\mathbf{P}_1$ as the PDA obtained directly from \textit{Construction A} in \cite{YCTC}. 

Next, we let $\mathbf{P}_1$ to be the  $(\Lambda,\binom{\Lambda}{t_1},\binom{\Lambda-1}{t_1-1},\binom{\Lambda}{t_1+1})$ MaN PDA for some $t_1\in [\Lambda]$. Also, we let $\mathbf{P}_2$ to be the $(\mathcal{L}_1,\binom{\mathcal{L}_1}{t_2},\binom{\mathcal{L}_1-1}{t_2-1},\binom{\mathcal{L}_1}{t_2+1})$ MaN PDA for some $t_2\in [\mathcal{L}_1]$. The $(K,\Lambda,\mathcal{L},F_{\text{MaN}},Z_{\text{MaN}},Z_{\text{MaN}}^{(h)},S_{\text{MaN}})$ SP-PDA $\mathbf{Q}_{\text{MaN}}$ obtained from $\mathbf{P}_1$ and $\mathbf{P}_2$ using our construction has the parameters $F_{\text{MaN}}=\binom{\Lambda}{t_1}\binom{\mathcal{L}_1}{t_2}, Z_{\text{MaN}}=\binom{\Lambda-1}{t_1-1}\binom{\mathcal{L}_1}{t_2}+\binom{\Lambda-1}{t_1}\binom{\mathcal{L}_1-1}{t_2-1},Z_{\text{MaN}}^{(h)}=\binom{\Lambda-1}{t_1-1}\binom{\mathcal{L}_1}{t_2}$, and $S_{\text{MaN}}=\sum_{n=1}^{\Lambda-t_1}\binom{\Lambda-n}{t_1}\left[\binom{\mathcal{L}_1}{t_2+1}-\binom{\mathcal{L}_1-\mathcal{L}_n}{t_2+1}\right]$. Notice that, this SP-PDA corresponds to \textit{Scheme 2} in \cite{PNR3} (see Remark \ref{remQfromMaNPDAs}). 

To have the number of helper caches to be the same in both the above cases, we let $\Lambda=q(m+1)$. Additionally, if we equate $M_h/N$ in both the cases, we get $1/q=t_1/\Lambda=t_1/(q(m+1))$. This implies $t_1=m+1$. Further, it is easy to verify that the normalized private cache size $M_p/N$ is the same in both the cases, where $M_p/N=(1-t_1/\Lambda)t_2/\mathcal{L}_1=(1-1/q)t_2/\mathcal{L}_1$. Now, we have
\begin{align*}
\frac{F_{\text{MaN}}}{F_A}& = \frac{\binom{\Lambda}{t_1}\binom{\mathcal{L}_1}{t_2}}{q^m\binom{\mathcal{L}_1}{t_2}}= \frac{\binom{\Lambda}{t_1}}{q^m}=\frac{\binom{\Lambda}{t_1}}{(\frac{\Lambda}{t_1})^{t_1-1}}\\&\approx \frac{t_1^{t_1-1}\Lambda^{t_1}}{\Lambda^{t_1-1}}=\Lambda t_1^{t_1-1}. 
\end{align*}  
Due to combinatorial expressions in $S_A$ and $S_{\text{MaN}}$, it is hard to compare the rates of the coded caching schemes corresponding to $\mathbf{Q}_A$ and $\mathbf{Q}_{\text{MaN}}$. However, we consider $\mathcal{L}$ to be uniform (each helper cache serves exactly $K/\Lambda$ users) and compare the rates. In that case, we have
\begin{align*}
R_A = \frac{S_A}{F_A} =\frac{ q^m(q-1)\binom{K/\Lambda}{t_2+1}}{q^m\binom{K/\Lambda}{t_2}}=\frac{(q-1)(K/\Lambda-t_2)}{t_2+1}.
\end{align*} 
Similarly, we have the rate expression corresponding to $\mathbf{Q}_{MaN}$ as follows:
\begin{align*}
R_{\text{MaN}} &= \frac{S_{\text{MaN}}}{F_{\text{MaN}}} = \frac{\binom{\Lambda}{t_1+1}\binom{K/\Lambda}{t_2+1}}{\binom{\Lambda}{t_1}\binom{K/\Lambda}{t_2}}\\&=\frac{(\Lambda-t_1)(K/\Lambda-t_2)}{(t_1+1)(t_2+1)}.
\end{align*} 
Therefore, we have
\begin{align*}
\frac{R_{\text{MaN}}}{R_A}&= \frac{\Lambda-t_1}{(q-1)(t_1+1)}=\frac{q(m+1)-(m+1)}{(q-1)(m+2)}\\	
&=\frac{m+1}{m+2}=\frac{t_1}{t_1+1}.
\end{align*}
Even though $\mathbf{Q}_A$ leads to a coded caching scheme with a slightly higher rate compared to \textit{Scheme 2} in \cite{PNR3}, there is a significant gain achieved in the subpacketization number (of the order of $\Lambda t_1^{t_1-1}$, $t_1\in [\Lambda]$).

\begin{figure*}
	\centering
	\begin{subfigure}{0.45\textwidth}
		\includegraphics[width=\textwidth]{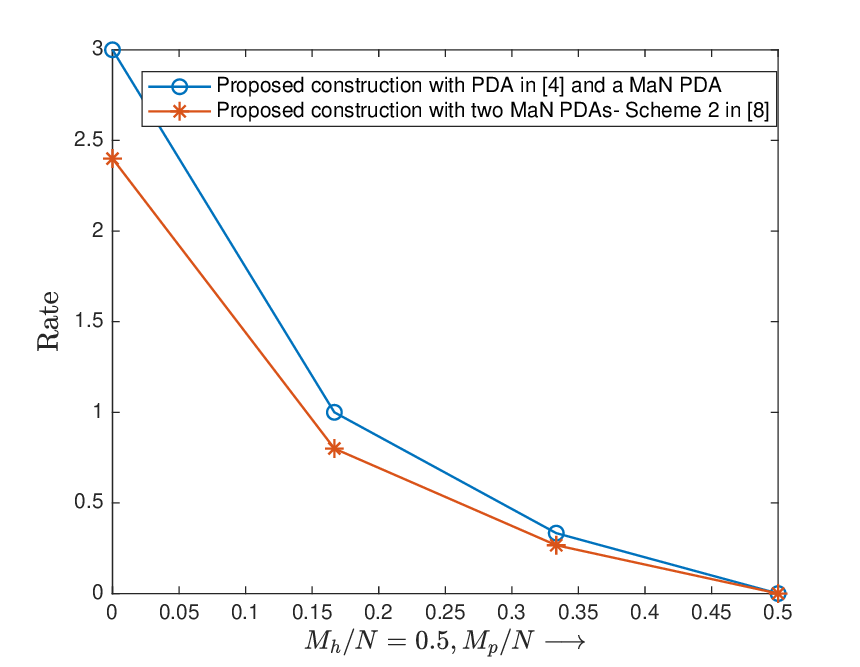}
		\caption{Rate against $M_p/N$}	
		\label{rateunif}
	\end{subfigure}
	\hfill
	\begin{subfigure}{0.45\textwidth}
		\includegraphics[width=\textwidth]{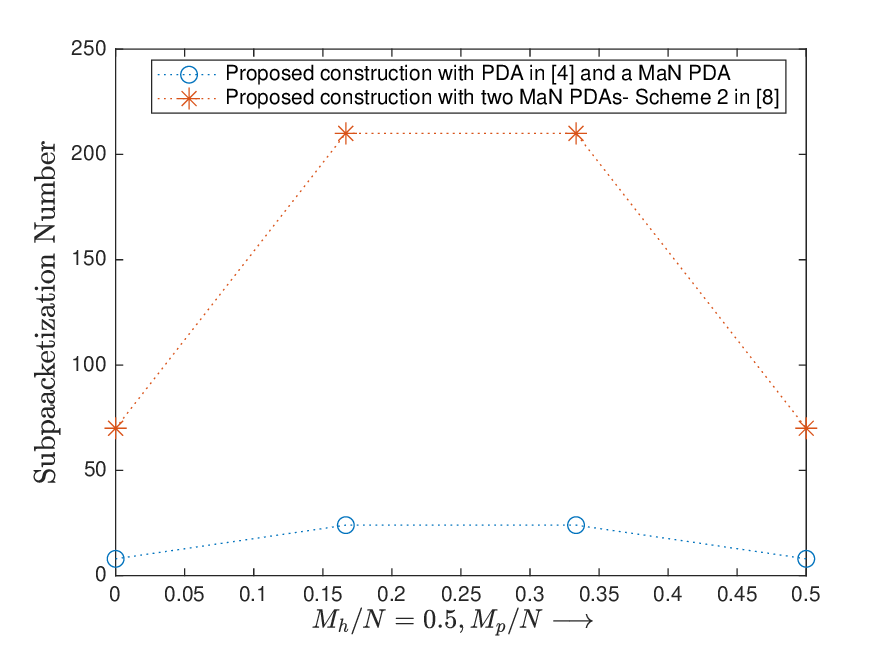}
		\caption{Subpacketization number against $M_p/N$}
		\label{subpacketunif}
	\end{subfigure}
	\caption{\mbox{$(K=24,\Lambda=8,\mathcal{L}=(3,3,\dots,3),M_h=N/2,M_p,N)$ coded caching scheme. }}
	\label{fig:unif}
\end{figure*}

\begin{figure*}
	\centering
	\begin{subfigure}{0.45\textwidth}
		\includegraphics[width=\textwidth]{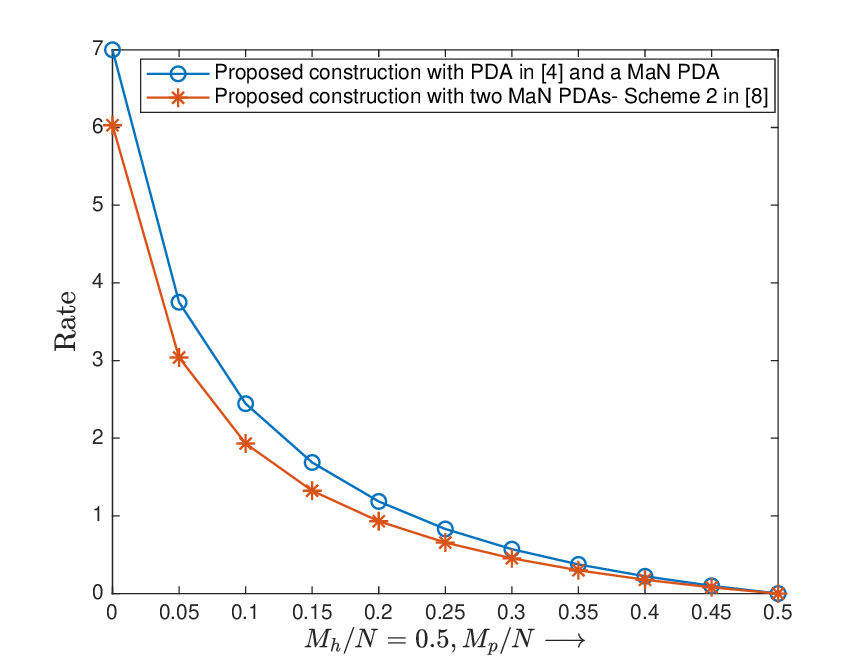}
		\caption{Rate against $M_p/N$}	
		\label{rateskewed}
	\end{subfigure}
	\hfill
	\begin{subfigure}{0.45\textwidth}
		\includegraphics[width=\textwidth]{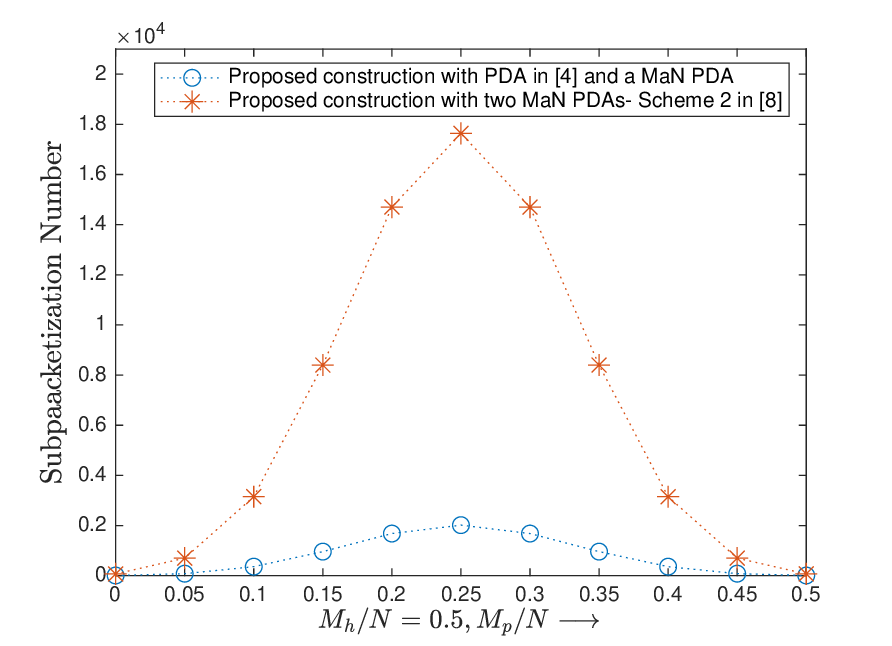}
		\caption{Subpacketization number against $M_p/N$}
		\label{subpacketskewed}
	\end{subfigure}
	\caption{\mbox{$(K=24,\Lambda=8,\mathcal{L}=(10,4,2,2,2,2,1,1),M_h=N/2,M_p,N)$ coded caching scheme. }}
	\label{fig:skewed}
\end{figure*}

We consider a coded caching network with $K=24,\Lambda=8,\mathcal{L}=(3,3,3,3,3,3,3,3)$, and $M_h/N=1/2$. In Fig. \ref{rateunif}, we plot the rates of the schemes obtained from $\mathbf{Q}_A$ and $\mathbf{Q}_{\text{MaN}}$ for different values of $M_p/N$. Further, in Fig. \ref{subpacketunif}, we show the corresponding subpacketization numbers. From the plots, it is clear that, we gain in subpacketization number by paying in rate. 

Next, we consider a coded caching network with $K=24,\Lambda=8,\mathcal{L}=(10,4,2,2,2,2,1,1)$, and $M_h/N=1/2$. In Fig. \ref{rateskewed}, we plot the rates of the schemes obtained from $\mathbf{Q}_A$ and $\mathbf{Q}_{\text{MaN}}$ for different values of $M_p/N$. Further, in Fig. \ref{subpacketskewed}, we show the corresponding subpacketization numbers. From the plots, it is evident that we gain in subpacketization number by a factor of around 10. However, the penalty in rate is minimal (approximately by a factor 7/6).
\section{Conclusion}
\label{Conclusion}
In this work, we introduced a combinatorial structure called SP-PDA to obtain coded caching schemes for a network where helper caches and private caches co-exist. The proposed SP-PDAs describe the helper cache placement, private cache placement, and the server transmissions in a single array. Further, we proposed a novel construction of SP-PDAs using two PDAs. Interestingly, we showed that the permutation of the columns of the two chosen PDAs will result in SP-PDAs with different performances. Then, we characterized the conditions for selecting the best column permutations of chosen PDAs. Furthermore, the coded caching schemes resulting from SP-PDAs subsume two coded caching schemes in \cite{PNR3} as special cases. Additionally, SP-PDAs enable the construction of coded caching schemes with much smaller subpacketization numbers compared to the existing schemes, without paying much penalty in terms of rate.  
\section*{Acknowledgment}
This work was supported partly by the Science and Engineering Research Board (SERB) of Department of Science and Technology (DST), Government of India, through J.C. Bose National Fellowship to Prof. B. Sundar Rajan, and by the Ministry of Human Resource Development (MHRD), Government of India, through Prime Minister’s Research Fellowship (PMRF) to K. K. Krishnan Namboodiri and Elizabath Peter.

\newpage
\begin{appendices}
	\section{Finding $S$ in an SP-PDA constructed from \\MaN PDAs }
	\label{app1}
	In this section, we find the number of integers in an SP-PDA $\mathbf{Q}$ obtained from the  $(\Lambda,\binom{\Lambda}{t_1},\binom{\Lambda-1}{t_1-1},\binom{\Lambda}{t_1+1})$ MaN PDA $\mathbf{P}_1$ and the $(\mathcal{L}_1,\binom{\mathcal{L}_1}{t_2},\binom{\mathcal{L}_1-1}{t_2-1},\binom{\mathcal{L}_1}{t_2+1})$ MaN PDA $\mathbf{P}_2$. From the proof of Theorem \ref{thm2}, we have the number of integers 
	\begin{equation*}
	S = \sum_{s=1}^{S_1}\phi_{\mathbf{P}_2}(\mathcal{L}_{\xi_{\mathbf{P}_1}(s)}).
	\end{equation*} 
	Consider the MaN PDA $\mathbf{P}_1$. The number of integers present in the first column of $\mathbf{P}_1$ is $\binom{\Lambda-1}{t_1}$. For those integers, we have 
	\begin{align*}
	\phi_{\mathbf{P}_2}(\mathcal{L}_{\xi_{\mathbf{P}_1}(s)}) = \phi_{\mathbf{P}_2}(\mathcal{L}_1)=\binom{\mathcal{L}_1}{t_2+1}.
	\end{align*}   
	In $\mathbf{P}_1$, the number of integers which are not present in the first column and present in the second column is $\binom{\Lambda-2}{t_1}$. For those integers, we have 
	\begin{align*}
	\phi_{\mathbf{P}_2}(\mathcal{L}_{\xi_{\mathbf{P}_1}(s)}) = \phi_{\mathbf{P}_2}(\mathcal{L}_2)=\binom{\mathcal{L}_1}{t_2+1}-\binom{\mathcal{L}_1-\mathcal{L}_2}{t_2+1}.
	\end{align*}     
	Similarly, in $\mathbf{P}_1$, the number of integers which are not present in the first $n$ columns and present in the $n^{\text{th}}$ column is $\binom{\Lambda-n}{t_1}$, where $n\in [\Lambda-t_1]$. For those integers, we have 
	\begin{align*}
	\phi_{\mathbf{P}_2}(\mathcal{L}_{\xi_{\mathbf{P}_1}(s)}) = \phi_{\mathbf{P}_2}(\mathcal{L}_n)=\binom{\mathcal{L}_1}{t_2+1}-\binom{\mathcal{L}_1-\mathcal{L}_n}{t_2+1}.
	\end{align*} 
	Since a number appears $t_1+1$ times in $\mathbf{P}_1$, no new integer will appear in the columns $\Lambda-t_1+1,\Lambda-t_1+2,\dots,\Lambda$ that are not present in the columns $1,2,\dots,\Lambda-t_1$. Therefore, we have the required expression
	\begin{equation*}
	S=\sum_{n=1}^{\Lambda-t_1}\binom{\Lambda-n}{t_1}\left[\binom{\mathcal{L}_1}{t_2+1}-\binom{\mathcal{L}_1-\mathcal{L}_n}{t_2+1}\right].
	\end{equation*}

		\section{Calculation of $S_A$ in Section \ref{compare}}
		\label{app2}
	Let $\mathbf{P}_1$ be the $(q(m+1),q^m,q^{m-1},q^m(q-1))$ PDA obtained from \textit{Construction A} in \cite{YCTC}. Also, let $\mathbf{P}_2$ be the $(\mathcal{L}_1,\binom{\mathcal{L}_1}{t_2},\binom{\mathcal{L}_1-1}{t_2-1},\binom{\mathcal{L}_1}{t_2+1})$ MaN PDA for some $t_2\in [\mathcal{L}_1]$. The $(K,\Lambda,\mathcal{L},F_A,Z_A,Z_A^{(h)},S_A)$ SP-PDA $\mathbf{Q}_A$ obtained from $\mathbf{P}_1$ and $\mathbf{P}_2$ using our construction. In this section, we show that $S_A\leq q^{m-1}(q-1)\left(\sum_{n=1}^{q} \binom{\mathcal{L}_1}{t_2+1}-\binom{\mathcal{L}_1-\mathcal{L}_n}{t_2+1}\right)$.
	
	 By the construction of $\mathbf{P}_1$ (\cite{YCTC}), we can partition the set of integers $[q^m(q-1)]$ into $q$ sets $\mathcal{S}_n,n\in [q]$, where $|\mathcal{S}_n|=q^{m-1}(q-1)$, for every $n\in [q]$. Also, we have $\xi_{\mathbf{P}_1}(s_1)=n$ for every $s_1\in \mathcal{S}_n$. For an integer $s_1\in \mathcal{S}_n$, we have 
	 	\begin{align*}
	 \phi_{\mathbf{P}_2}(\mathcal{L}_{\xi_{\mathbf{P}_1}(s_1)}) = \phi_{\mathbf{P}_2}(\mathcal{L}_n)=\binom{\mathcal{L}_1}{t_2+1}-\binom{\mathcal{L}_1-\mathcal{L}_n}{t_2+1}
	 \end{align*} 
	since $\mathbf{P}_2$ is the $(\mathcal{L}_1,\binom{\mathcal{L}_1}{t_2},\binom{\mathcal{L}_1-1}{t_2-1},\binom{\mathcal{L}_1}{t_2+1})$ MaN PDA. Therefore, we have the required expression
	\begin{equation*}
	S_A=q^{m-1}(q-1)\sum_{n=1}^{q}\left[\binom{\mathcal{L}_1}{t_2+1}-\binom{\mathcal{L}_1-\mathcal{L}_n}{t_2+1}\right].
	\end{equation*}
	
	\end{appendices}

\end{document}